\documentclass[
aps,
prd,
nofootinbib,
twocolumn,
superscriptaddress,
floatfix
]{revtex4-2}%采用revtex4-1有doi没title
\usepackage{amsmath}
\allowdisplaybreaks
\usepackage{graphicx}
\usepackage{longtable}
\usepackage{multirow}
\usepackage{epsfig}
\usepackage[usenames]{color}
\usepackage[
colorlinks=true,
linkcolor=blue,
citecolor=blue,
urlcolor=blue
]{hyperref}
\begin{document}
%\begin{CJK*}{GBK}{song}
%\begin{CJK*}{GB}{}

\title{Shear viscosity and electric conductivity of quark matter at finite temperature and chemical potential with QCD phase transitions }

\author{Wei-bo He}
\affiliation{ MOE Key Laboratory for Non-equilibrium Synthesis and Modulation of Condensed Matter, 
	School of Physics, Xi’an Jiaotong University, Xi’an 710049, China}
\author{Guo-yun Shao}   
\email[ ]{gyshao@mail.xjtu.edu.cn} 
%%\thanks{These authors contributed equally to this work}
\affiliation{ MOE Key Laboratory for Non-equilibrium Synthesis and Modulation of Condensed Matter, 
School of Physics, Xi’an Jiaotong University, Xi’an 710049, China}
\author{Chong-long Xie}
\affiliation{ MOE Key Laboratory for Non-equilibrium Synthesis and Modulation of Condensed Matter, School of Physics, Xi’an Jiaotong University, Xi’an 710049, China}

\author{Ren-xin Xu} 
%\email[ ]{r.x.xu@pku.edu.cn}
\affiliation{ School of Physics, Peking University, Beijing, 100871, China}

\begin{abstract}

In the Beam Energy Scan phase II (BES-II) experiments at RHIC STAR, the quark-gluon plasma (QGP) produced with changing collision energies may probe different regions of the QCD phase diagram. Correspondingly, studying the transport coefficients of quark matter in these regions will contribute to extracting the QCD phase structure through hydrodynamic approaches. We investigate the shear viscosity and electric conductivity within the framework of kinetic theory with the relaxation time approximation, in particular their behaviors  near the Mott and first-order phase transitions with a spinodal structure as well as along the isentropic trajectories. To derived the scattering cross-section under different conditions, the temperature and chemical potential dependent masses of quarks, antiquarks and exchanged mesons are calculated in the Polyakov-loop extended Nambu--Jona Lasinio~(PNJL) model. The numerical results indicate that, at small chemical potential, the shear viscosity to entropy density ratio ($\eta/s$) has a minimum near the Mott phase transition and increases rapidly in the lower-temperature side of the chiral crossover phase transition. At large chemical potential (high baryon density), $\eta/s$ in the QGP phase is dominated by temperature, and the value of $\eta/s$ is greatly enhanced at low temperatures. At intermediate temperature and chemical potential near the QCD phase transition, the behavior of $\eta/s$ is influenced by the competition between temperature, density, and QCD phase transition. 
The electirc conductivity ($\sigma/T$) roughly exhibits similar characteristics to $\eta/s$ in the QCD phase diagram,
whereas the dimensionless ratio of $\eta/s$ to $\sigma/T$ decreases monotonically with growing temperature, approaching a constant in the high-temperature limit.
\end{abstract}

%\pacs{12.38.Mh, 25.75.Nq}

\maketitle
\section{introduction}

Understanding the nature of quark-gluon plasma  and the phase transition to hadronic matter is currently one of the most important questions in quantum chromodynamics~(QCD). The calculation from lattice QCD and hadron resonance gas model indicates that the QCD phase transition is a smooth crossover at small chemical potential~\cite{aoki2006order, gupta2011scale, bazavov2012chiral, borsanyi2013freeze,bazavov2014equation, bazavov2017skewness,borsanyi2014full,borsanyi2020qcd}. The simulation of lattice QCD is limited at large baryon chemical potential due to the fermion-sign problem.
Many QCD inspired approaches/models predict that a first-order phase transition with a second-order critical end point (CEP) exists at finite temperature and chemical potential, connecting with the crossover transformation~(e.g., \cite{fukushima2008phase, ratti2006phases, meisinger1996chiral, costa2010phase, sasaki2012theta, ferreira2014deconfinement, schaefer2010thermodynamics, skokov2011quark, qin2011phase,gao2016phase, fischer2014phase, maslov2023effect, fu2020qcd, shao2018baryon, he2022speed, liu2021three}).

The theoretical description of the dynamical evolution of QGP in heavy-ion collisions is needed to build the connection with experimental data. The relativistic viscous hydrodynamics provides an effective framework to describe the collective behaviors and transport properties of QGP~\cite{israel1979transient, baier2008relativistic, denicol2012derivation, jeon2015introduction}. The shear viscosity ($\eta$) and bulk  viscosity ($\zeta$) are parameters that quantify  the dissipation processes in the evolution of fluid. They reflect the deviation of a medium from an ideal fluid. The simulations indicate that the final hadronic observables are sensitive to the two viscous parameters~\cite{romatschke2007viscosity, song2008suppression, denicol2009effect}. 

In the earlier studies, the ratios of shear and bulk viscosity to entropy density  $\eta/s$ and $\zeta/s$ were taken as constants to simulate the evolution of QGP in high-energy collisions~\cite{heinz2013collective, gale2013hydrodynamic}.
One crucial achievement from the combination of heavy-ion collisions and the viscous hydrodynamics was the
discovery that the QGP is a nearly perfect fluid~\cite{teaney2003effect, shuryak2004toward,shuryak2005rhic} with the shear viscosity to entropy density ratio $\eta/s$ at very small chemical potential close to $1/{4\pi}$, the KSS limit derived in
anti–de Sitter space/conformal field theory (AdS/CFT) correspondence~\cite{kovtun2005viscosity, policastro2001shear}.
Subsequently, it was argued that the value of  $\eta/s$ is sensitive to the phase transitions of QGP medium.
Similar to water, helium and nitrogen,  the shear viscosity to  density ratio $\eta/s$ of QGP possibly exhibits a minimum at the
transition from hadrons to quarks and gluons~\cite{csernai2006strongly,lacey2007has}. Since the direct calculation of these transport coefficients from the non-perturbative QCD  on lattice is difficult~\cite{kovtun2005viscosity, astrakhantsev2017temperature}, a growing number of studies were performed in effective models%[10–17]
, such as the hadron resonance gas (HRG) model~\cite{noronha2009transport, tiwari2012description, noronha2012hadron, kadam2015dissipative},  color string percolation model~\cite{sahu2021thermodynamic}, linear sigma model~\cite{heffernan2020hadronic}, quasiparticle models~\cite{mitra2017transport, moreau2019exploring,soloveva2020transport, mykhaylova2019quark,mykhaylova2021impact}, Green–Kubo
formalism~\cite{plumari2012shear,kharzeev2008bulk,harutyunyan2017bulk,czajka2017kubo,gubser2008thermodynamics,li2015temperature}, transport approach~\cite{wesp2011calculation, ozvenchuk2013shear, rose2018shear},  and QCD motivated models~\cite{haas2014gluon, christiansen2015transport, dubla2018towards,ghiglieri2018qcd,danhoni2023hot, gao2018temperature,zhang2018shear,chen2010shear,chen2013shear,gao2018temperature,sasaki2010transport,marty2013transport,soloveva2021shear,deb2016estimating,kadam2021critical,reichert2021first,deng2023shear,mclaughlin2022building, Fodor2018,rehberg1996elastic, xiao2014bulk}. 
The Bayesian statistical analysis has also been adopted to constrain the temperature dependent behavior of $\eta/s$ and $\zeta/s$ at vanishing or small chemical potential~\cite{bernhard2019bayesian, bernhard2016applying, everett2021multisystem,nijs2021bayesian, shen2023viscosities}.

In experiments, the BES II program at RHIC STAR provides a great opportunity to explore the evolution 
of QGP in a wide range of temperature and baryon chemical potential~\cite{luo2017search,STAR2025zdq,STAR2025owm}. Especially, the critical region and the first-order phase transition  may be involved in the BES II experiments with the collision energy declined to several GeV.  As the shear viscosity is  closely correlated with the collective flows of final state particles, its value are crucial to diagnose the QCD phase transition in future hydrodynamical simulations of QGP evolution with a first-order phase transition. Although some calculations have been performed to explore the shear viscosity of hot and dense quark matter~\cite{marty2013transport,soloveva2021shear,deb2016estimating,soloveva2020transport,shen2023viscosities,  deng2023shear,mclaughlin2022building,rehberg1996elastic}, even including the chiral phase transition in the temperature-baryon chemical ($T-\mu_B$) plane~\cite{marty2013transport,soloveva2021shear,mclaughlin2022building},
an issue that needs further investigation
is the behavior of shear viscosity of quark matter near the Mott phase transition and the first-order chiral phase transition associated with a spinodal structure as well as the evolution of  $\eta/s$  on the isentropic trajectories. On the other hand, the electric conductivity ($\sigma$) is the parameter to quantify the Ohmic response of QGP to an electric field. It is of fundamental importance for the evolution and lifetime of the electromagnetic fields in the QGP medium~\cite{mitra2017transport}.
In this work, we will elaborate on the shear viscosity and electric conductivity of quark matter in the full QCD phase diagram. The behaviors of $\eta/s$ and $\sigma/T$  on different phase planes ($T-\mu_B$ and $T-\rho_B$ planes, as well as along the isentropic trajectories) will be explored in detail, particularly their behaviors near the meson Mott dissociation and the chiral phase transition, including the crossover, critical and first-order phase transition with the  spinodal structure connected with the metstable and unstable phase. We will also calculate the ratio of $\eta/s$ to $\sigma/T$, and analyze its characteristics in high-temperature limit and near the QCD phase transitions.

The calculations will be performed within the kinetic theory with the relaxation time approximation based on the $2 \rightarrow 2$ elastic scatterings between  $u,d, s$ quarks and their antiparticles. To derive the cross section and flavor dependent relaxation time, the medium dependent masses of quasiparticles will be calculated in the PNJL model. The paper is organized as follows. In Sec.~II, we introduce the 2+1 flavor PNJL quark model and the formulas to calculate the shear viscosity and electric conductivity in kinetic theory with the relaxation time approximation. In Sec.~III, we illustrate the numerical results of cross section, flavor-dependent relaxation time, shear viscosity and electric conductivity under different physical coditions, and discuss their relations with the QCD phase transitions. A summary is given in Sec. IV.

\section{PNJL quark model and formulae of shear viscosity and electric conductivity }
\subsection{The 2+1 flavor PNJL quark model}
We employ the 2+1 flavor PNJL to describe the properties of QGP medium. 
The Lagrangian density is written as
\begin{eqnarray}\label{L}
	\!\mathcal{L}&\!=&\!\bar{\psi}(i\gamma^{\mu}\!D_{\mu}\!+\!\gamma_0\hat{\mu}\!-\!\hat{m}_{0})\psi\!+\!
	G\!\sum_{k=0}^{8}\!\big[(\bar{\psi}\lambda_{k}\psi)^{2}\!+\!
	(\bar{\psi}i\gamma_{5}\lambda_{k}\psi)^{2}\big]\nonumber \\
	&&-K\big[\texttt{det}_{f}(\bar{\psi}(1+\gamma_{5})\psi)+\texttt{det}_{f}
	(\bar{\psi}(1-\gamma_{5})\psi)\big]\nonumber \\ 
	&&-U(\Phi[A],\bar{\Phi}[A],T),
\end{eqnarray}
where $\psi$ denotes the quark fields with three flavors, $u,\ d$, and
$s$; $\hat{m}_{0}=\texttt{diag}(m_{u},\ m_{d},\
m_{s})$ in flavor space; $G$ and $K$ are the four-point and
six-point interacting constants, respectively.  The $\hat{\mu}=diag(\mu_u,\mu_d,\mu_s)$ are the quark chemical potentials.
The covariant derivative is defined as $D_\mu=\partial_\mu-iA_\mu$.
The gluon background field $A_\mu=\delta_\mu^0A_0$ is supposed to be homogeneous
and static, with  $A_0=g\mathcal{A}_0^\alpha \frac{\lambda^\alpha}{2}$, where
$\frac{\lambda^\alpha}{2}$ is $SU(3)$ color generators.

The effective potential $U(\Phi[A],\bar{\Phi}[A],T)$ is expressed with the traced Polyakov loop
$\Phi=(\mathrm{Tr}_c L)/N_c$ and its conjugate
$\bar{\Phi}=(\mathrm{Tr}_c L^\dag)/N_c$. The Polyakov loop $L$  is a matrix in color space
\begin{equation}
	L(\vec{x})=\mathcal{P} exp\bigg[i\int_0^\beta d\tau A_4 (\vec{x},\tau)   \bigg],
\end{equation}
where $\beta=1/T$ is the inverse of temperature and $A_4=iA_0$. 
The Polyakov-loop effective potential \cite{roessner2007polyakov} taken in this study  is
\begin{eqnarray} \label{U}
	\frac{U(\Phi,\bar{\Phi},T)}{T^4}&=&-\frac{a(T)}{2}\bar{\Phi}\Phi +b(T)\,\mathrm{ln}\big[1-6\bar{\Phi}\Phi\\ \nonumber
	&&+4(\bar{\Phi}^3+\Phi^3)-3(\bar{\Phi}\Phi)^2\big],
\end{eqnarray}
where
\begin{equation}
	\!a(T)\!=\!a_0\!+\!a_1\big(\frac{T_0}{T}\big)\!+\!a_2\big(\frac{T_0}{T}\big)^2 \,\,\,\texttt{and}\,\,\,\,\, b(T)\!=\!b_3\big(\frac{T_0}{T}\big)^3,
\end{equation}
with  $a_0=3.51$,  $a_1=-2.47$,  $a_3=15.2$, $b_3=-1.75$ and $T_0=210\,$ MeV.
The physical quark mass can be derived in the mean field approximation as
\begin{equation}
	M_{i}=m_{i}-4G\phi_i+2K\phi_j\phi_k\ \ \ \ \ \ (i\neq j\neq k),
	\label{mass}
\end{equation}
where $\phi_i$ stands for the quark condensate of the flavor $i$.

The mesons are constructed with the quark-antiquark effective interaction. The pseudoscalar mesons $(\pi^0, \pi^{ \pm}, K^0, \overline{K^0}, K^{ \pm}, \eta, \eta^{\prime})$, and the scalar meson partners $\left(\sigma_{\pi^0}, \sigma_{\pi^{ \pm}}, \sigma_{K^0}, \sigma_{\overline{K^0}}, \sigma_{K^{ \pm}}, \sigma, \sigma^{\prime}\right)$ are considered to describe the elastic scatterings between quarks and antiquarks.  In Eq.~(\ref{L})   the six-fermion interaction in the determinant term 
can be reduced to an effective four-point
interaction in the mean-field approximation, by contracting out $\bar{\psi}\psi$ pairs. 
We currently consider the $SU(2)$ isospin symmetry for $u$ and $d$ quarks with the degeneracy condition $m_u=m_d$ and  $\mu_u=\mu_d$, the effective Lagrangian can be then derived as

\begin{eqnarray}
	\mathcal{L}= && \bar{\psi}(i\gamma^{\mu}D_{\mu}\!+\!\gamma_0\hat{\mu}\!-\!\hat{m}_{0})\psi  -U(\Phi[A], \bar{\Phi}[A], T)\nonumber \\
	&&+\sum_{a=0}^8\left[K_a^{-}\left(\bar{\psi} \lambda^a \psi\right)^2+\right. \left.K_a^{+}\left(\bar{\psi} i \gamma_5 \lambda^a \psi\right)^2\right] \\ 
	&&+K_{08}^{-}\left[\left(\bar{\psi} \lambda^8 \psi\right)\left(\bar{\psi} \lambda^0 \psi\right)+\left(\bar{\psi} \lambda^0 \psi\right)\left(\bar{\psi} \lambda^8 \psi\right)\right] \nonumber \\
	&&+K_{08}^{+}[\left(\bar{\psi} i \gamma_5 \lambda^8 \psi\right)\left(\bar{\psi} i \gamma_5 \lambda^0 \psi\right)+\left(\bar{\psi} i \gamma_5 \lambda^0 \psi\right)\left(\bar{\psi} i \gamma_5 \lambda^8 \psi\right)].\nonumber 
\end{eqnarray}

 The coupling constants in above equation fulfills the following relations
\begin{equation}
	\begin{aligned}
		K_{0}^{\pm}&=G\mp\frac{1}{3}K({\cal G}^{u}+{\cal G}^{d}+{\cal G}^{s}),\\ 
		K_{1}^{\pm}&=K_{2}^{\pm}=K_{3}^{\pm}=G\pm\frac{1}{2}K{\cal G}^{s},\\
		K_{4}^{\pm}&=K_{5}^{\pm}=K_{6}^{\pm}=K_{7}^{\pm}=G\pm\frac{1}{2}K{\cal G}^{u},\\
		K_{8}^{\pm}&=G\pm\frac{1}{6}K(2{\cal G}^{u}+2{\cal G}^{d}-{\cal G}^{s}),\\
		K_{08}^{\pm}&=\pm\frac{1}{12}\sqrt{2}K({\cal G}^{u}+{\cal G}^{d}-{2\cal G}^{s}),
	\end{aligned}
\end{equation}
where
\begin{equation}
	\mathcal{G}^f=-\frac{N_c}{4\pi^2} A_f(T,\mu),
\end{equation}
and
\begin{equation}
	A_f(T, \mu)=16 \pi^2 \int_0^{\Lambda} \frac{d^3 p}{(2 \pi)^3}\left[1- \frac{n_f^{+}}{2 E}\right.\left.- \frac{n_f^{-}}{2 E}\right].
\end{equation}
The $n_f^{+}$ and $n_f^{-}$ are the modified effective Fermion distribution functions of quark and antiquark with
\begin{equation}\label{distribution}
	\!\!\!n^+_f\!=\!\frac{\!\Phi e^{\!-\!(\!E_i\!-\!\mu_i\!)\!/\!T}\!+\!2\bar{\Phi} e^{\!-\!2(\!E_i\!-\!\mu_i\!)\!/\!T}+e^{-3(E_i-\mu_i)/T}}
	{\!1\!+\!3\Phi e^{\!-\!(\!E_i\!-\!\mu_i\!)\!/\!T}\!+\!3\bar{\Phi} e^{\!-\!2(\!E_i\!-\!\mu_i)/T}\!+\!e^{\!-\!3(E_i\!-\!\mu_i)\!/\!T}}
\end{equation}
and
\begin{equation}\label{antif}
	\!\!\bar n_f^{-}\!=\!\frac{\! \bar \Phi e^{\!-\!(\!E_i\!+\!\mu_i\!)\!/\!T}\!+\!2\Phi e^{\!-\!2(\!E_i\!+\!\mu_i\!)\!/\!T}+e^{-3(E_i+\mu_i)/T}}
	{\!1\!+\!3\bar \Phi e^{\!-\!(\!E_i\!+\!\mu_i\!)\!/\!T}\!+\!3\Phi e^{\!-\!2(\!E_i\!+\!\mu_i\!)/T}\!+\!e^{\!-\!3(\!E_i\!+\!\mu_i)\!/\!T}} ,
\end{equation}
where $E_{i}=\sqrt{\mathbf{p}^{2}+M_{i}^{2}}$ is the dispersion relation of (anti)quarks in QCD medium.

With the isospin symmetry of $u$ and $d$ quark, the $\pi^0$ is decoupled from the $\eta$ and $\eta^{\prime}$, and the $\pi^{ \pm}, \pi^0$ are degenerate. Similarly, $K_4^{ \pm}=K_6^{ \pm}$, which means that the neutral and charged kaons have the same mass.  The quark-quark scattering amplitude will be calculated in the
random phase approximation~\cite{rehberg1996elastic}, which leads to
\begin{equation}\label{pion}
	\mathcal{D}_{\pi(\sigma_\pi)}=\frac{2 K_3^{ \pm}}{1-4 K_3^{ \pm} \Pi_{q \bar{q}}^{P(S)}\left(p_0, \vec{p}\right)}
\end{equation}
for the pion~($\sigma_\pi$) propagator and 
\begin{equation}\label{kaon}
	\mathcal{D}_{K(\sigma_K)}=\frac{2 K_4^{ \pm}}{1-4 K_4^{ \pm} \Pi_{q \bar{s}}^{P(S)}\left(p_0, \vec{p}\right)}
\end{equation}
for the kaon~($\sigma_K$) propagator. The label $q$ stands for $u$ and $d$ quark in Eqs.~(\ref{pion})-(\ref{kaon}).

In the rest system of reference, the pion mass $M_\pi$ and the kaon mass  $M_K$ satisfy the following relations
\begin{equation}\label{pionm}
	1-4 K_3^{+} \Pi_{u \bar{u}}^P\left(M_\pi-i \frac{\Gamma_\pi}{2}, 0\right)=0
\end{equation}
and
\begin{equation}\label{kaonm}
	1-4 K_4^{+} \Pi_{u \bar{s}}^P\left(M_K-i \frac{\Gamma_K}{2}, 0\right)=0,
\end{equation}
where $\Gamma_\pi$ and $\Gamma_K$ are the decay width of pion and kaon, respectively.
With the increase of temperature, when the value of $\Gamma_\pi$ or $\Gamma_K$ changes from zero to nonzero, the dissociation from mesons to quarks occurs, which is also called the Mott phase transition. The polarization function
$\Pi_{f \bar{f}^{\prime}}^{P(S)}\left(p_0,\vec{p}\right)$ derived in Ref. \cite{rehberg1996elastic} is 
\begin{eqnarray}
		\Pi_{f \bar{f}^{\prime}}^{P(S)}\left(p_0, \mathbf{k}\right)\! =&&\!\!\!\!-\frac{N_c}{8 \pi^2}\left\{A\left(M_f, \mu_f, T\right)+A\left(M_{f^{\prime}}, \mu_{f^{\prime}}, T\right)\right. \nonumber\\
		&& \!+\!\left[\left(M_f \mp M_{f^{\prime}}\right)^2\!-\!\left(p_0+\mu_f-\mu_{f^{\prime}}\right)^2\!+\!\mathbf{p}^2\right] \nonumber \\
		&& \left.\times B_0\left(|\mathbf{p}|, M_f, \mu_f, M_{f^{\prime}}, \mu_{f^{\prime}}, p_0, T\right)\right\},
\end{eqnarray}
where $B_0\left(|\mathbf{p}|, M_f, \mu_f, M_{f^{\prime}}, \mu_{f^{\prime}}, p_0, T\right)$ is the source of the imaginary part of the polarization function and the meson mass. It can be written as
\begin{widetext}
\begin{eqnarray}
		 B_0\left(|\mathbf{p}|, M_f, \mu_f, M_{f^{\prime}}, \mu_{f^{\prime}}, p_0, T\right)=&& \tilde{B}_0^{+}\left(-\lambda,|\mathbf{p}|, M_f, M_{f^{\prime}}, \mu_f, \beta, \varphi\right) 
		 -\tilde{B}_0^{-}\left(\lambda,|\mathbf{p}|, M_f, M_{f^{\prime}}, \mu_f, \beta, \varphi\right) \nonumber\\
		&& +\tilde{B}_0^{+}\left(\lambda,|\mathbf{p}|, M_{f^{\prime}}, M_f, \mu_{f^{\prime}}, \beta, \varphi\right) 
		 -\tilde{B}_0^{-}\left(-\lambda,|\mathbf{p}|, M_{f^{\prime}}, M_f, \mu_{f^{\prime}}, \beta, \varphi\right),
\end{eqnarray}
where
\begin{equation}
		 \tilde{B}_0^{ \pm}\left(\lambda,|\mathbf{p}|, M_1, M_2, \mu_1, \mu_2, T, \varphi\right)=  2 \int_{M_1}^{\sqrt{M_1^2+\Lambda^2}}  \texttt{d} E\,\,  k\,n^{ \pm}_f \,   \int_{-1}^{+1} \frac{1}{\lambda^2+2 \lambda E+2 k|\mathbf{p}| x-|\mathbf{p}|^2+M_1^2-M_2^2} \text{d} x,
\end{equation}
\end{widetext}
in which $\lambda=p_0+\mu_1-\mu_2$ and $E=\sqrt{k^2+M_1^2}$. 

For the $\eta$ and $\eta^{\prime}$, the quark-antiquark scattering matrix involves the mixing term
\begin{equation}
	\mathcal{D}_{\eta, \eta^{\prime}}=2 \frac{\mathcal{K}}{\mathcal{AC} - \mathcal{B}^2}\left(\begin{array}{ll}
		\mathcal{A} & \mathcal{B} \\
		\mathcal{B} & \mathcal{C}
	\end{array}\right)
\end{equation}
with 
\begin{equation}
	\begin{aligned}
		& \mathcal{A}=K_0^{+}-\frac{4}{3} \mathcal{K}\left(\Pi_{u \bar{u}}^P+2 \Pi_{s \bar{s}}^P\right) \\
		& \mathcal{B}=K_{08}^{+}+\frac{4}{3} \sqrt{2} \mathcal{K}\left(\Pi_{u \bar{u}}^P-\Pi_{s \bar{s}}^P\right) \\
		& \mathcal{C}=K_8^{+}-\frac{4}{3} \mathcal{K}\left(2 \Pi_{u \bar{u}}^P+\Pi_{s \bar{s}}^P\right) \\
		& \mathcal{K}=K_0^{+} K_8^{+}-K_{08}^2
	\end{aligned}
\end{equation}

In the numerical calculation, a cut-off $\Lambda$ is implemented in the three-momentum
space for divergent integrations. We take the model parameters obtained in~\cite{rehberg1996hadronization}:
$\Lambda=602.3$ MeV, $G\Lambda^{2}=1.835$, $K\Lambda^{5}=12.36$,
$m_{u,d}=5.5$  and $m_{s}=140.7$ MeV, determined
by fitting $f_{\pi}=92.4$ MeV,  $M_{\pi}=135.0$ MeV, $m_{K}=497.7$ MeV and $m_{\eta}=957.8$ MeV. 

\subsection{Shear viscosity and electric conductivity in kinetic theroy with the relaxation time approximation}

We shall calculate the transport coefficients in the PNJL model based on the kinetic theory in which quarks, antiquarks and mesons are taken as quasiparticle. The shear viscosity and electric conductivity from the kinetic theory with relaxation time approximation\cite{hosoya1985transport,gavin1985transport,sasaki2009bulk,sasaki2010transport,chakraborty2011quasiparticle,albright2016quasiparticle,soloveva2020transport} are written as
\begin{equation}\label{sv}
	\eta=\frac{1}{15 T} \sum_{\alpha} d_{q}\tau_\alpha \int \frac{d^3 p}{(2 \pi)^3} \frac{p^4}{E_\alpha^2} f_{\alpha}^{0}(1-f_{\alpha}^{0})
\end{equation}
\begin{equation}
	\sigma_{e} = \frac{e^{2}}{3T}
	\sum_{\alpha}q_{\alpha}^{2}d_{q}\tau_{\alpha}
	\int\frac{d^{3}p}{(2\pi)^3}
	\frac{p^2}{E_{\alpha}^{2}}f_{\alpha}^{0}(1-f_{\alpha}^{0})
\end{equation}
where the subscript $\alpha$ stands for quarks~($u,d,s$) and antiquarks~($\bar{u},\bar{d},\bar{s}$), $d_q=2N_c$ is the degeneracy factor for spin and color, $e^{2}=4\pi/137$ is the scaled charge-squared with quark charges $q_{i}=+2/3(u),-1/3(d),-1/3(s)$.
The relaxation time $\tau_\alpha$ for particles $\alpha$ depends on the environment of QGP medium, i.e., it varies with temperature and chemical potential. $f^0_\alpha\left(E_\alpha, T, \mu_\alpha\right)$ describes the distribution function in local equilibrium for quarks and antiquarks. The effective distribution function given in Eqs.~(\ref{distribution}) and ~(\ref{antif}) will be used in the calculation. 

Eq.~(\ref*{sv}) indicates that the relaxation time plays a very important role in calculating the shear viscosity coefficient. It is related to the microscopic scattering process between  particles  in the thermal medium. In this work, the $2 \rightarrow 2$ elastic scatterings, including the quark-quark, quark-antiquark and antiquark-antiquark, are considered to calculate the shear viscosity of QGP. The averaged relaxation time of particle species $i$ can be derived as
\begin{equation}
	\tau_i^{-1}\left(T, \mu_q\right)=\sum_{j=q, \bar{q}} \rho_j\left(T, \mu_q\right) \bar{w}_{i j},
\end{equation}
where $\rho_j\left(T, \mu_q\right)$ is the quark or antiquark number density of species $j$. $ \bar{w}_{i j}$ is the  averaged transition rate defined as
\begin{equation}\label{wij}
	\begin{aligned}
		\bar{w}_{i j}= & \frac{1}{\rho_i \rho_j} \int \frac{d^3 \mathbf{p}_i}{(2 \pi)^3} \int \frac{d^3 \mathbf{p}_j}{(2 \pi)^3} d_q f_i^{(0)}\left(E_i, T, \mu_q\right) d_q \\
		& \times f_j^{(0)}\left(E_j, T, \mu_q\right) \cdot v_{\mathrm{rel}} \cdot \sigma_{i j \rightarrow c d}\left(s, T, \mu_q\right) .
	\end{aligned}
\end{equation}
Here $v_{\mathrm{rel}}$ is the relative velocity and $\sqrt{s}$ is the center-of-mass energy. $\sigma_{i j \rightarrow c d}$ is the cross section from  the initial incident (anti)quarks $(i, j)$ to  the final outgoing (anti)quarks $(c, d)$ at collision energy $\sqrt{s}$.

The relative velocity $v_{\mathrm{rel}}$  can be calculated in the c.m frame with \cite{2012Eric,soloveva2021shear}
\begin{equation}
	v_{\text {rel }}  =\frac{\sqrt{\left(E_i^* E_j^*-\mathbf{p}_{i}^{*}\cdot\mathbf{p}_{j}^{*}\right)^2-\left(M_i M_j\right)^2}}{E_i^* E_j^*}
\end{equation}
where
\begin{equation}
	E_i^*  =\frac{s+M_i^2-M_j^2}{2 \sqrt{s}},\,\,\,\,
E_j^*  =\frac{s-M_i^2+M_j^2}{2 \sqrt{s}} 
\end{equation}
and 
\begin{equation}
	\left|\mathbf{p}_i^*\right| =\frac{\sqrt{\left(s-\left(M_i+M_j\right)^2\right) \cdot\left(s-\left(M_i-M_j\right)^2\right)}}{2 \sqrt{s}} .
\end{equation}

The total cross section can be derived with the integral of the differential cross section
\begin{equation}\label{cs}
	\sigma=\int_{t_{-}}^{t^{+}} \mathrm{d} t \frac{\mathrm{d} \sigma}{\mathrm{d} t}\left(1-f^{(0)}(E_c^*, T, \mu_q)\right)\left(1-f^{(0)}(E_d^*, T, \mu_q)\right)
\end{equation}
where  the factor $(1-f^{(0)})$ represents the Pauli blocking for quarks and/or antiquarks in the final state; $t$ is the Mandelstam variable; $t_{\pm}$ indicates the upper and lower limits of $t$ \cite{soloveva2021shear}
\begin{equation}
	\begin{aligned}
		t_{ \pm}= & M_i^2+M_c^2-\frac{1}{2 s}\left(s+M_i^2-M_j^2\right)\left(s+M_c^2-M_d^2\right) \\
		&\!\pm \! 2\!\left.\sqrt{\!\frac{\left(s\!+\!M_i^2\!-\!M_j^2\right)^2}{4 s}\!-\!M_i^2}\right.\! \sqrt{\frac{\left(s\!+\!M_c^2\!-\!M_d^2\right)^2}{4 s}\!-\!M_c^2} .
	\end{aligned}
\end{equation}
The differential cross section in Eq.~(\ref{cs}) is written as 
\begin{equation}
	\frac{d \sigma}{d t}=\frac{1}{16 \pi s_{i j}^{+} s_{i j}^{-}} \frac{1}{4 N_c^2} \sum_{s c}|\mathcal{M}|^2,
\end{equation}
where the  $s_{ij}^\pm$ is defined as
\begin{equation}
	s_{i j}^{ \pm}=s-\left(M_i \pm M_j\right)^2. 
\end{equation}
 $ \frac{1}{4 N_c^2}\sum_{s c}|{\mathcal{M}}|^2$ denotes the matrix element squared averaged over the color and spin of the incident particles, and summed over the final scattered particles. The calculation of  $|{\mathcal{M}}|^2$ is related to the specific scattering channels, which can be performed using the standard methods in quantum field theory. One can refer to Refs.~\cite{rehberg1996elastic,soloveva2021shear} for more details.

For the quark-quark scattering, only $u$ and $t$ channel are involved. The mesons exchanged in different scattering processes are listed in Table.~\ref{T1}.  Due to the isospin symmetry of $u, d$ quark, the process of $u + s \rightarrow u + s$ is equivalent to that of $d + s \rightarrow d + s$, and the process of $u + u \rightarrow u + u$ is equivalent to that of $d + d \rightarrow d + d$. For the quark-antiquark scattering, only $t$ and $s$ channel are involved. The mesons exchanged in different processes are listed in Table.~\ref{T2}. The antiquark-antiquark scattering can be calculated using the same steps as the quark-quark scattering.
\begin{table}[h]
	\caption{Mesons exchanged in the $t$ and $u$ channels in the different quark-quark scattering processes.}
	\label{T1}
\begin{tabular}{lcc}
	\hline \hline process & \begin{tabular}{c} 
		exchanged mesons \\
		 ($u$ channel)
	\end{tabular} & \begin{tabular}{c} 
		exchanged mesons \\
		 ($t$ channel)
	\end{tabular} \\
	\hline$u d \rightarrow u d$ & $\pi, \sigma_\pi$ & $\pi, \eta, \eta^{\prime}, \sigma_\pi, \sigma, \sigma^{\prime}$ \\
	$u u \rightarrow u u$ & $\pi, \eta, \eta^{\prime}, \sigma_\pi, \sigma, \sigma^{\prime}$ & $\pi, \eta, \eta^{\prime}, \sigma_\pi, \sigma, \sigma^{\prime}$ \\
	$u s \rightarrow u s$ & $K, \sigma_K$ & $\eta, \eta^{\prime}, \sigma, \sigma^{\prime}$ \\
	$s s \rightarrow s s$ & $\eta, \eta^{\prime}, \sigma, \sigma^{\prime}$ & $\eta, \eta^{\prime}, \sigma, \sigma^{\prime}$ \\
	\hline \hline
\end{tabular}
\end{table}

\begin{table}[h]
	\caption{Mesons exchanged in the $s$ and $t$ channels for in the different quark-antiquark scattering processes.}
	\label{T2}
	\begin{tabular} {lcc}
		\hline \hline Process & \begin{tabular}{c} 
			exchanged mesons   \\
			($s$ channel)
		\end{tabular} & \begin{tabular}{c} 
			exchanged mesons \\
			( $t$ channel)
		\end{tabular}\\
		\hline$u \bar{d} \rightarrow u \bar{d}$ & $\pi, \sigma_\pi$ & $\pi, \eta, \eta^{\prime}, \sigma_\pi, \sigma, \sigma^{\prime}$ \\
		$u \bar{u} \rightarrow u \bar{u}$ & $\pi, \eta, \eta^{\prime}, \sigma_\pi, \sigma, \sigma^{\prime}$ & $\pi, \eta, \eta^{\prime}, \sigma_\pi, \sigma, \sigma^{\prime}$ \\
		$u \bar{u} \rightarrow d \bar{d}$ & $\pi, \eta, \eta^{\prime}, \sigma_\pi, \sigma, \sigma^{\prime}$ & $\pi, \sigma_\pi$ \\
		$u \bar{s} \rightarrow u \bar{s}$ & $K, \sigma_K$ & $\eta, \eta^{\prime}, \sigma, \sigma^{\prime}$ \\
		$u \bar{u} \rightarrow s \bar{s}$ & $\eta, \eta^{\prime}, \sigma, \sigma^{\prime}$ & $K, \sigma_K$ \\
		$s \bar{s} \rightarrow u \bar{u}$ & $\eta, \eta^{\prime}, \sigma, \sigma^{\prime}$ & $K, \sigma_K$ \\
		$s \bar{s} \rightarrow s \bar{s}$ & $\eta, \eta^{\prime}, \sigma, \sigma^{\prime}$ & $\eta, \eta^{\prime}, \sigma, \sigma^{\prime}$ \\
		\hline \hline
	\end{tabular}
\end{table}

\section{Numerical results and discussion}

In this section, we will present the numerical results, including the meson Mott dissociation, the cross section, the flavor dependent relaxation time and the shear viscosity and electric conductivity across the QCD phase diagram.  In particular, we will discuss how the relaxation processes and transport coefficients depend on temperature, chemical potential (net baryon density), and the various phase transitions.

\subsection{Mott dissociation of pseudoscalar mesons}
The dissociation of mesons is closely related to the microscopic scattering of particles in QCD medium. It involves the physical mass and constitute mass of meson in the medium. When the mass of a bound meson equals the constituent one, the meson becomes unstable and undergoes a Mott dissociation~(Mott phase transition)~\cite{blaschke2017mott}. The corresponding temperature is referred to as the Mott phase transition temperature, $T_{\text{Mott}}$. Above the  $T_{\text{Mott}}$, mesons will dissociate into quark-antiquark pairs. From the onset of Mott phase transition the meson decay width becomes non-zero and increases with rising temperature. 

\begin{figure}[htbp]
	\begin{center}
		\includegraphics[scale=0.37]{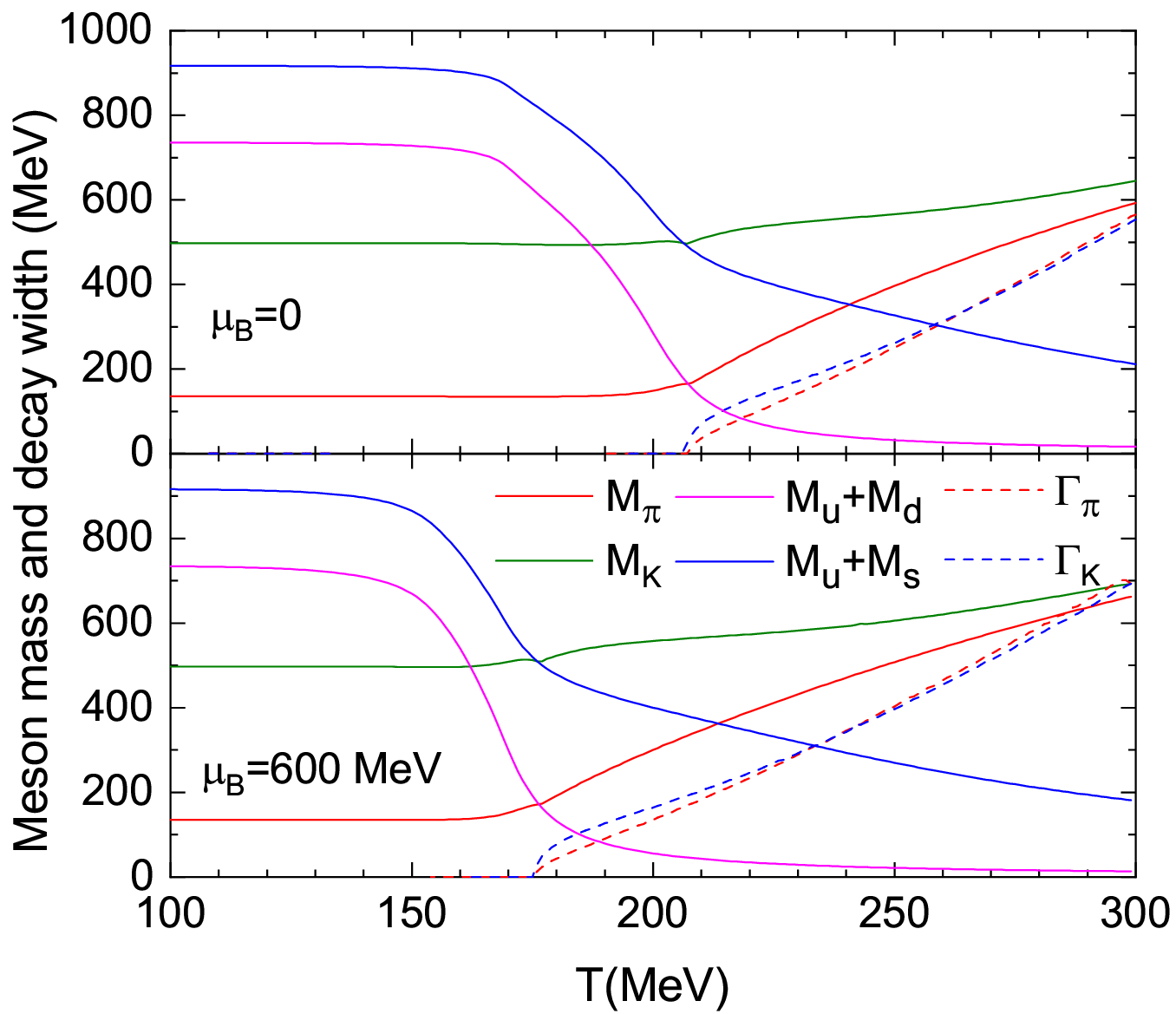}
		\caption{\label{fig:1} Masses of pseudoscalar mesons~(pion and kaon), the constitute masses $M_u+M_d$ and $M_u+M_s$, and the meson decay widths as functions of temperature for $\mu_B=0$ (upper) and  $\mu_B=600$~MeV (lower).
		}
	\end{center}
\end{figure}

Figure~\ref{fig:1} presents the physical masses of pseudoscalar mesons~(pion and kaon), the constitute masses  and the decay width as functions of temperature at  $\mu_B=0$ (upper) and  $\mu_B=600\,$MeV (lower). It shows that the constitute mass $M_u+M_d$~($M_u+M_s$) of pion~(kaon) decreases with increasing temperature due to the partial restoration of chiral symmetry. However, the physical mass of bound pion~(kaon) derived from Eqs.~(\ref{pionm}) and (\ref{kaonm}) increases with the rising temperature. Fig.~\ref{fig:1} also indicates that  the Mott phase transition at $\mu_B=600$~MeV is lower than that at $\mu_B=0$. This is because that not only the temperature but also the density effect can cause the restoration of chrial symmetry.  
It can be also observed that the pion and kaon undergo the Mott dissociations at nearly the same but slightly different temperature.

\begin{figure}[htbp]
	\begin{center}
		\includegraphics[scale=0.37]{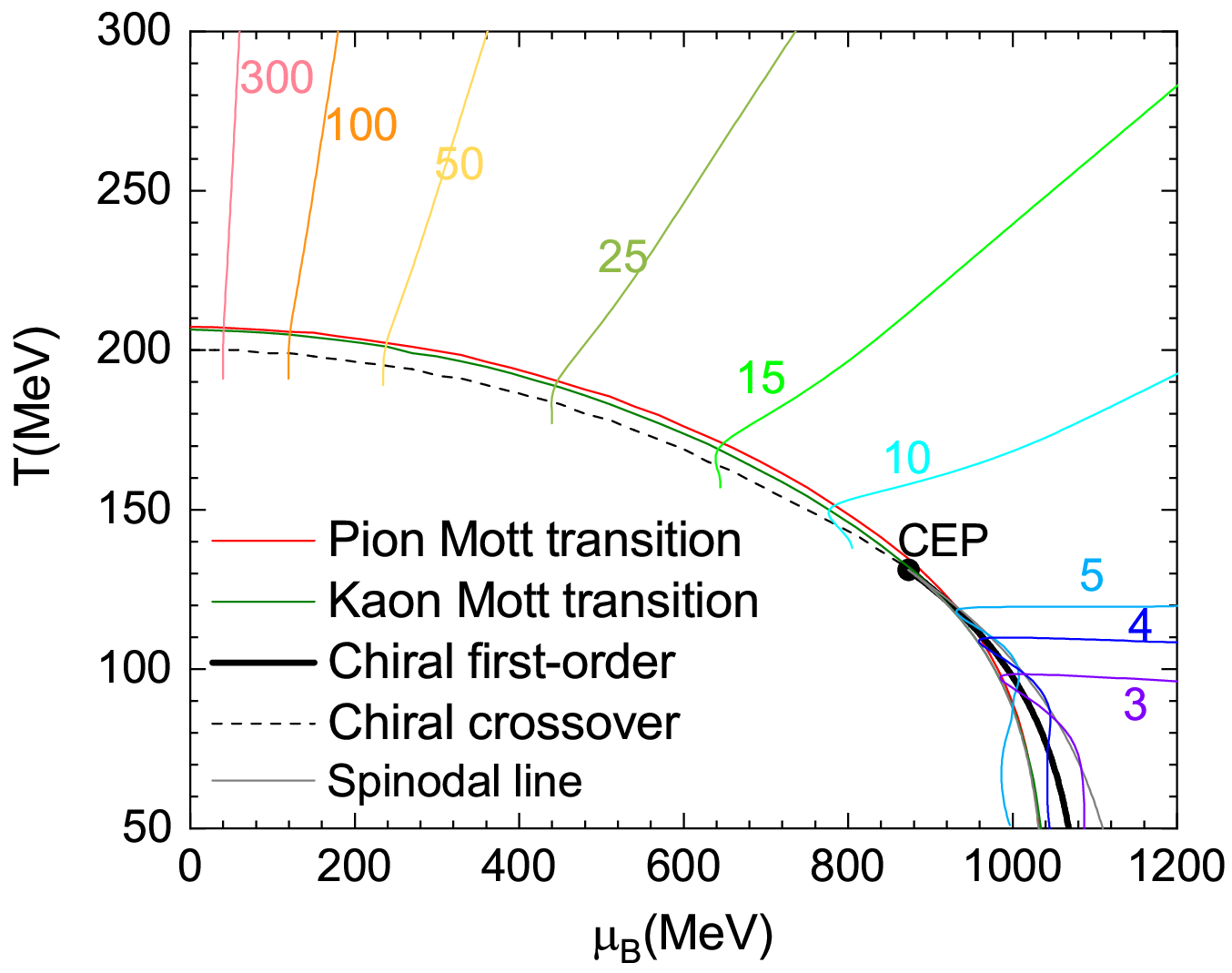}
		\caption{\label{fig:2} QCD phase diagram in the $T-\mu_B$ plane in the PNJL model, including the Mott phase transitions of pion and kaon and chiral crossover and first-order phase transitions with spinodal lines. The CEP locates $T_C=131\,$MeV, $\mu_C=873\,$MeV. The isentropic trajectories for $s/\rho_B=300,100,50,25,15,10,5,4,3$ are also plotted for the convenience of subsequent discussions.
		}
	\end{center}
\end{figure}

\begin{figure}[htbp]
	\begin{center}
		\includegraphics[scale=0.37]{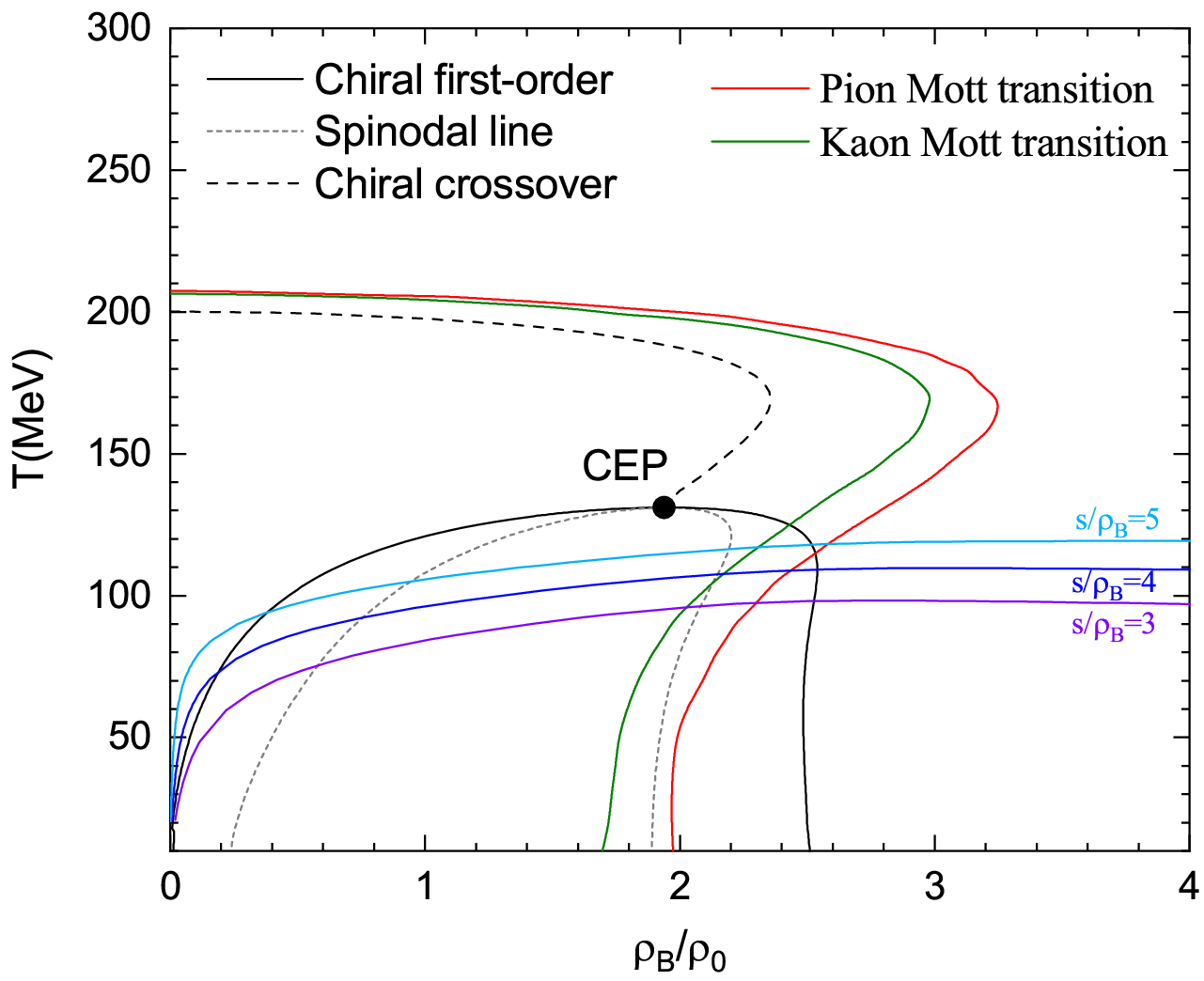}
		\caption{\label{fig:3} QCD phase diagram in the $T-\rho_B$ plane in the PNJL model, including the Mott phase transitions of pion and kaon and chiral crossover and first-order phase transitions with spinodal lines. The CEP locates $T_C=131\,$MeV, $\rho_C=1.94\rho_0$). The isentropic trajectories for $s/\rho_B=5,4,3$ are also illustrated.
		}
	\end{center}
\end{figure}

To indicate the relationship between the Mott phase transition of pseudoscalar mesons and the quark chiral phase transition, we plot in Figs.~\ref{fig:2} and \ref{fig:3} these phase transition lines in the full QCD phase diagram.
Fig.~\ref{fig:2} demonstrates that for each given chemical potential the pion and kaon Mott phase transition temperatures are close to but slightly higher than that of chiral crossover transition of $u$~($d$) quark. The difference between them is less than $10$ MeV in the whole crossover region, and continually decreases when approaching the CEP. Although these phase transition lines are close to each other in the $T-\mu_B$ plane near the critical region, there is a relatively larger difference  in the temperature-density plane, as demonstrated in Fig.~\ref{fig:3}. This is attributed to that the net baryon density in the critical region being sensitive to the chemical potential.


For a first-order phase transition, it is always accompanied by the presence of unstable and metastable phases in thermodynamics.
As shown in Fig.~\ref{fig:3}, the spinodal line represents the boundary of the unstable state. Within the spinodal line, the compressibility is negative with the mechanical instability of system. The binodal line (first-order phase transition line) is the boundary between metastable and stable states. The region between the spinodal line and the binodal line is the metastable phase. For a given temperature, the baryon chemical potential $\mu_B$ exhibits a non-monotonic dependence on density across the first-order phase transition—increasing, then decreasing, and finally increasing again—as the system passes through the metastable and unstable regions. The spinodal line is defined by the locus of points where the derivative $\partial \mu_B/\partial \rho_B$ vanishes \cite{Shao2020djd}. Fig.~\ref{fig:3} also indicates that the Mott dissociation  occurs in the metastable phase for pion and both the metastable and unstable  phases for kaon in the first-order phase transition range at  lower temperatures.

For smaller values of $s/\rho_B$, for example, $s/\rho_B=3,4,5$, the isentropic trajectories cross the first-order phase transition line as well as the metastable and unstable regions.
The corresponding spinodal sturcture and isentropic trajectories in the $T-\mu_B$ plane are also plotted in Fig.~\ref{fig:2}, which predicates the underlying complexity of the transport properties of QGP in the evolution near the first-order phase transition when the spinodal region is involved.

\subsection{Flavor dependent relaxation time}

\begin{figure}[htbp]
	\begin{center}
		\includegraphics[scale=0.31]{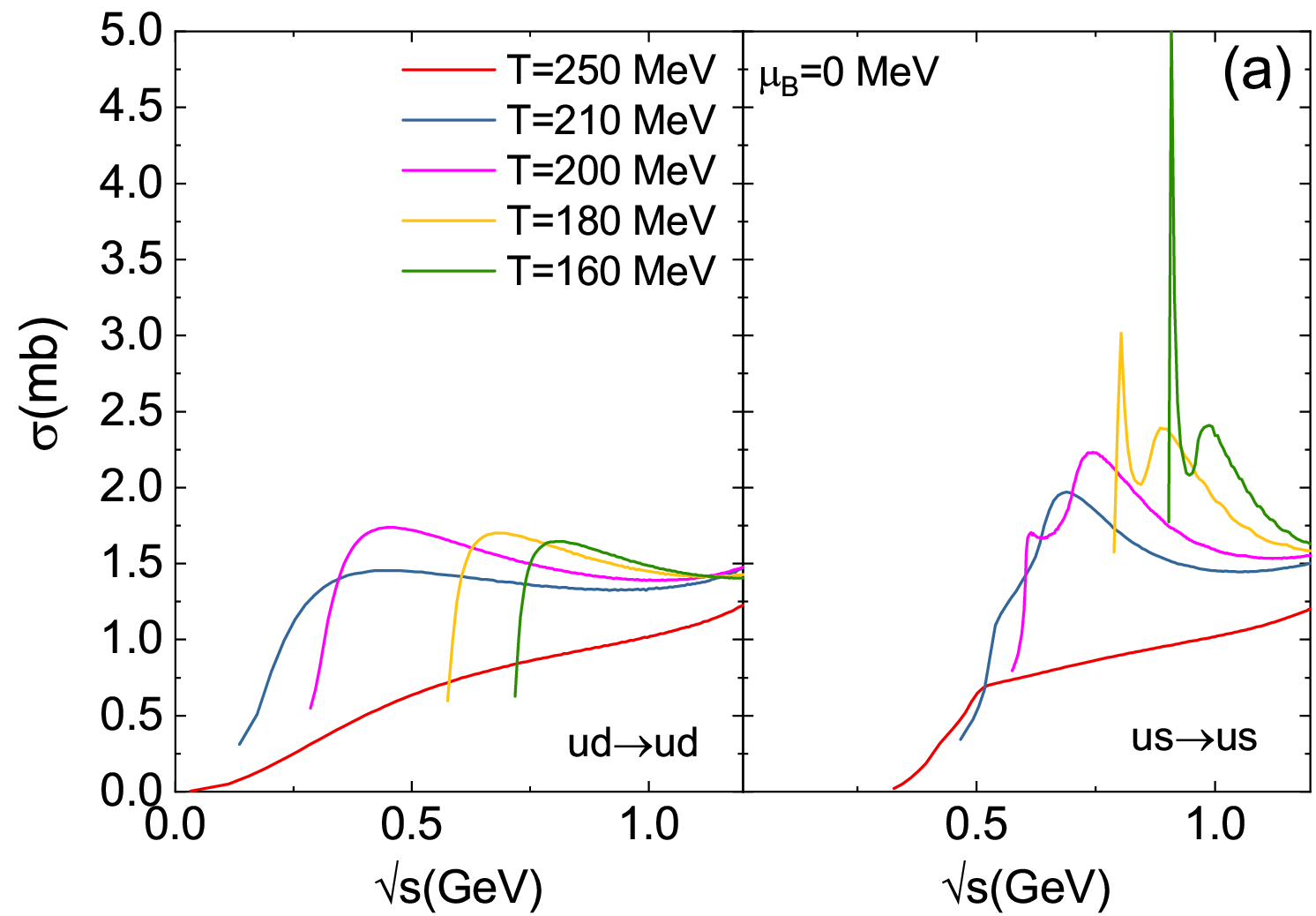}
		\includegraphics[scale=0.31]{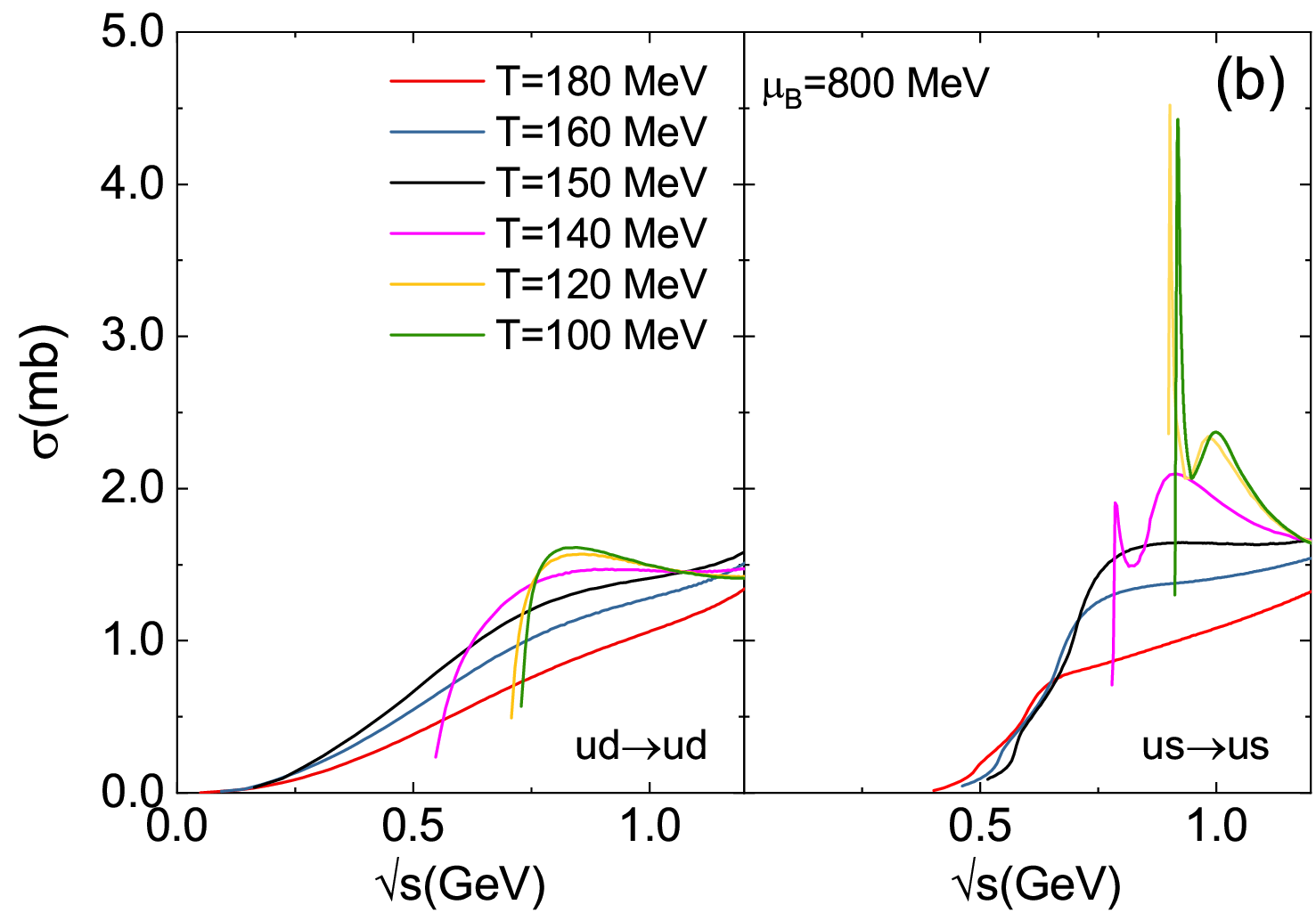}
		\includegraphics[scale=0.31]{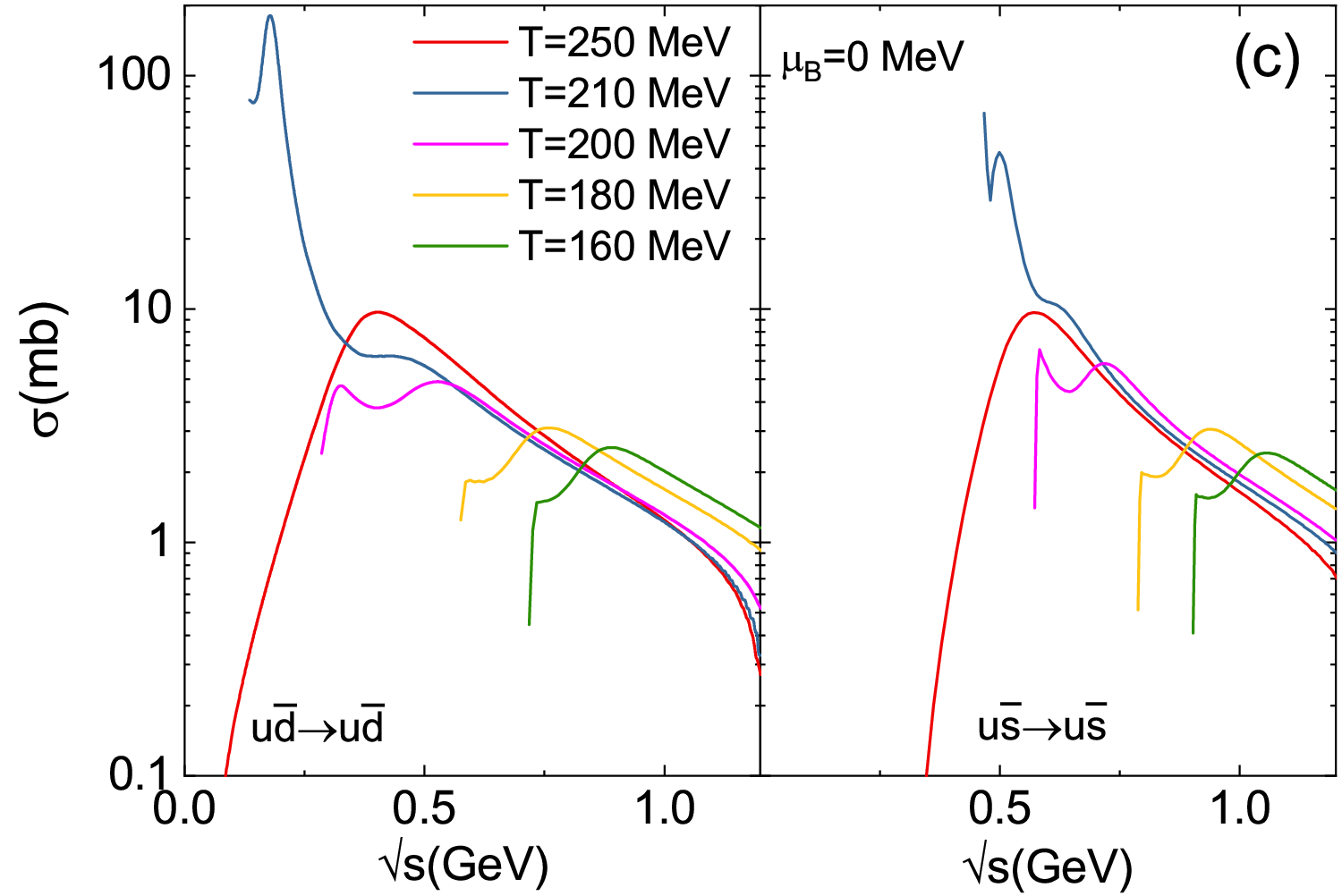}
		\includegraphics[scale=0.31]{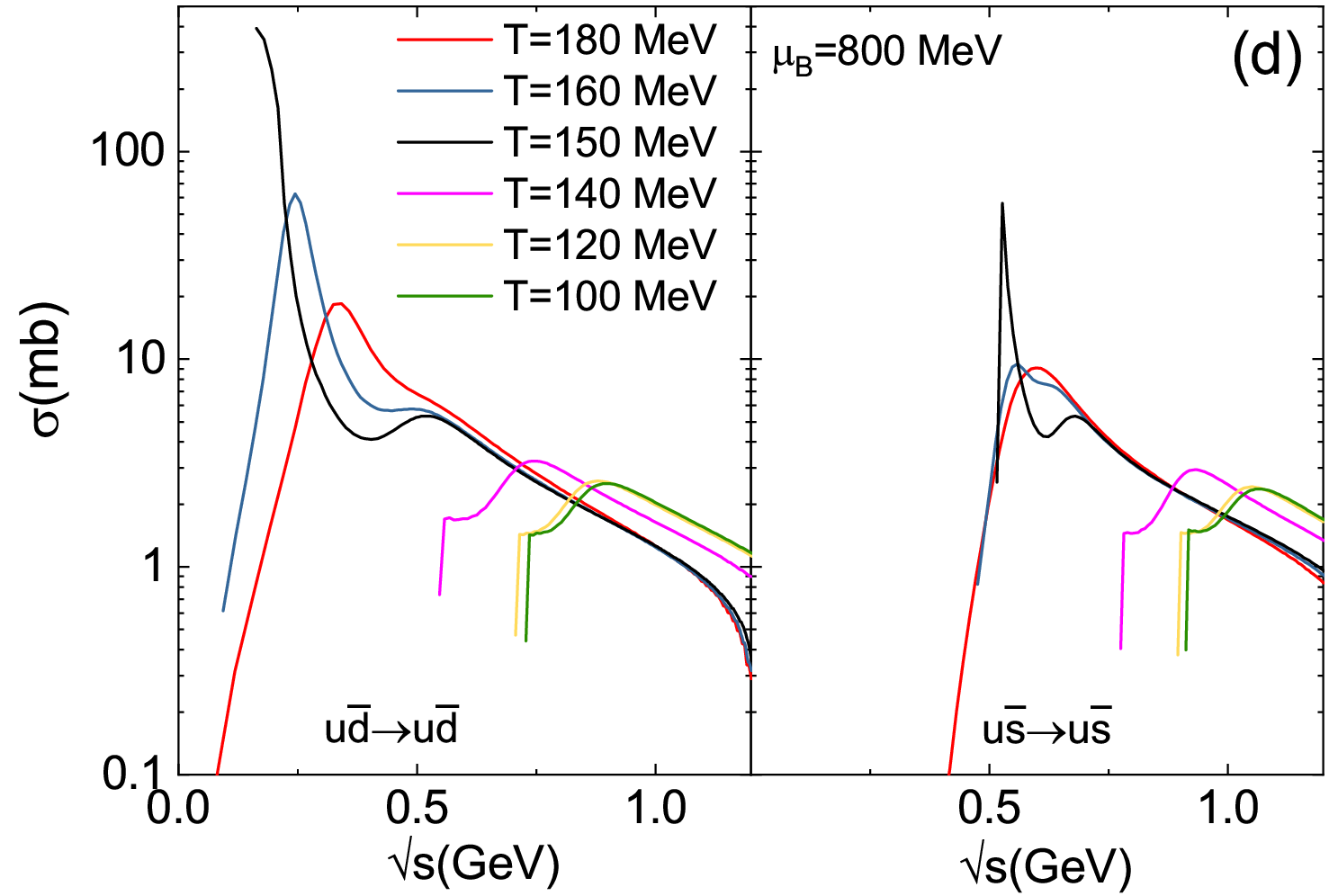}
		\caption{\label{fig:4} Cross sections of quark-quark (  $ud\rightarrow ud$ and $us\rightarrow us$) and quark-antiquark ( $u\bar{d}\rightarrow u\bar{d}$ and $u\bar{s}\rightarrow u\bar{s}$) elastic scattering as functions of center-of-mass collision energy at different temperature and quark chemical potential.
		}
	\end{center}
\end{figure}

The relaxation time of quark and antiquark plays a crucial role on the transport properties of QCD medium in kinetic theory with relaxation time approximation, which is related to the cross sections  among different particles.

We present in Fig.~\ref{fig:4} the cross sections of the $2\rightarrow2$ elastic scattering for quark-quark and quark-antiquark scatterings as functions of center-of-mass energy $\sqrt{s}$ at $\mu_B=0$ and $\mu_B=800$ MeV. For each scattering process, it is found that the cross section is relatively larger at the temperatures near the crossover phase transition and Mott phase transition.
%For the quark-quark scattering, Fig.~\ref{fig:4}(a) and~(b) present that 
When the temperature is much higher than that of Mott phase transition, the cross sections decrease since the $\pi$ and $K$ mesons have larger decay widths, as shown in Fig.~\ref{fig:1}.
Fig.~\ref{fig:4} also shows that the thresholds of particles in each process change with temperature and chemical potential. Compared with the $ud\rightarrow ud$~($u\bar{d}\rightarrow u\bar{d}$) scattering, the threshold of $us\rightarrow us$ ( $u\bar{s}\rightarrow u\bar{s}$) process at the same temperature and chemical potential  is postponed to a larger $\sqrt{s}$. The reason is that $\sqrt{s}$ should at least equal to the larger of the sum of incident particle mass and the sum of outgoing particle mass, and $M_u+M_s$ is always larger than $M_u+M_d$ in the whole QCD phase diagram.
Fig.~\ref{fig:4} also displays that some peak values in quark-antiquark scattering are obviously lager than those in quark-quark scattering. In the $u\bar{d}{\rightarrow}u\bar{d}$ and $u\bar{s}{\rightarrow}u\bar{s}$ scattering, this phenomenon appears at $\mu_{B}=0$ MeV,  $T=180$, $210$ and $250\,$MeV and $\mu_{B}=800\,$MeV,  $T=150$, $160$ and $180\,$MeV.
This is mainly attributed to the resonance of the exchanged meson with the incident quarks in the $s$ channel in quark-antiquark scattering, which leads to a distinct peak-like structure existing in Fig.~\ref{fig:4}(c) and~(d).

\begin{figure}[htbp]
	\centering
	\includegraphics[scale=0.37]{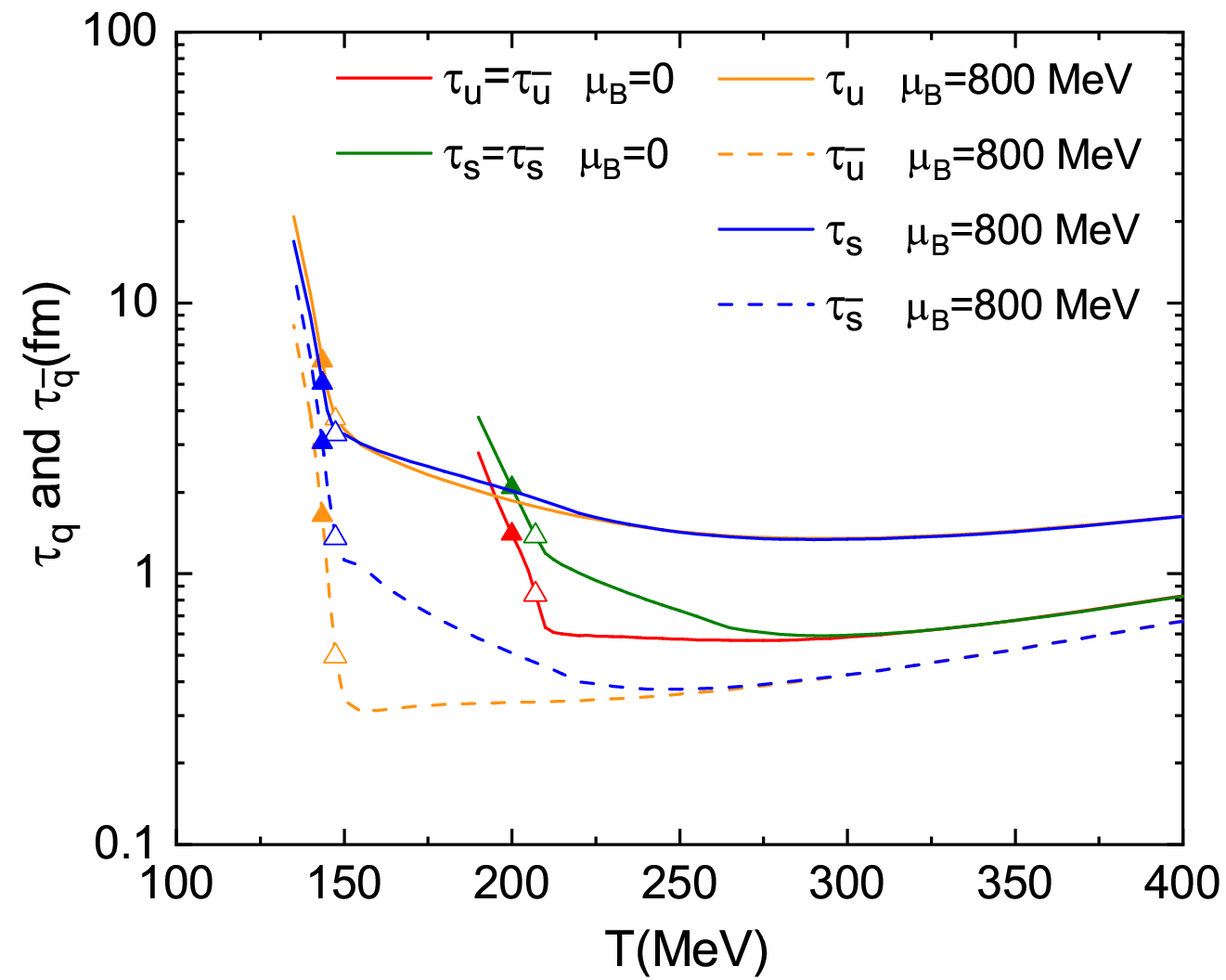}
	\caption{Relaxation time of quarks and antiquarks as functions of temperature for $\mu_B=0$ and $\mu_B=800\,$MeV. 
	The solid (hollow) triangles represent the $\eta/s$ at the point where these  $\mu_B(T)$ line intersect the chiral crossover phase transition line (pion Mott transition line).	}
		\label{fig:5} 
\end{figure}

Fig.~\ref{fig:5} presents the relaxation time of different particle species as functions of temperature at  $\mu_B=0$ and $\mu_B=800$ MeV. For the $u, d$ isospin symmetric quark matter, the relaxation time $\tau_u=\tau_d$ and $\tau_{\bar {u}}=\tau_{\bar{d}}$  always hold. In particular,  $\tau_u=\tau_d=\tau_{\bar{u}}=\tau_{\bar{ d}}$ at vanishing chemical potential.  In the case  $\mu_B=0$, it can be seen that the relaxation time $\tau_s$ of strange quark is larger than that of $u$~($d$) quark near the region of chiral and mott phase transitions. The chiral crossover transition temperature of $s$ quark is about $250$ MeV at $\mu_B=0$. We can see that  $\tau_s$ gradually approaches  $\tau_u$ at high temperature after the chiral restoration of $s$ quark, which is just the expected behavior under the SU(3) symmetry.
At finite chemical potential, this figure shows that $\tau_u \neq \tau_{\bar{u}}$ and $\tau_s \neq \tau_{\bar{s}}$  due to the different distribution function of quark and antiquark.
Numerically, the relaxation time of quarks are larger than those of antiquarks. 

Fig.~\ref{fig:5} also indicates that the relaxation for each particle species has a  concave shape, which is to certain degree connected with the shear viscosity of QGP medium.
The relaxation time describes the time scale required for a system to return to equilibrium after being subjected to a small perturbation. A shorter relaxation time indicates stronger interactions and more frequent collisions between particles. The shear viscosity coefficient is a macroscopic hydrodynamic quantity that describes a fluid's ability to resist shear deformation, i.e., the relative motion between different fluid layers. A smaller viscosity coefficient means the fluid flows more ``smoothly". From the perspective of kinetic theory, the shear viscosity coefficient is proportional to the relaxation time. A short relaxation time implies frequent collisions between particles, leading to localized momentum that cannot be effectively transferred between fluid layers. Consequently, the fluid is more prone to shear deformation, meaning the viscosity coefficient is small. Therefore, in the phase diagram where the relaxation time has a  concave shape, the viscosity coefficient probably also exhibit a minimum. In the next subsection we will see that the numerical results indeed exhibit this trend.

\subsection{Shear viscosity under different conditions }

\begin{figure}[htbp]
	\centering
	\includegraphics[scale=0.37]{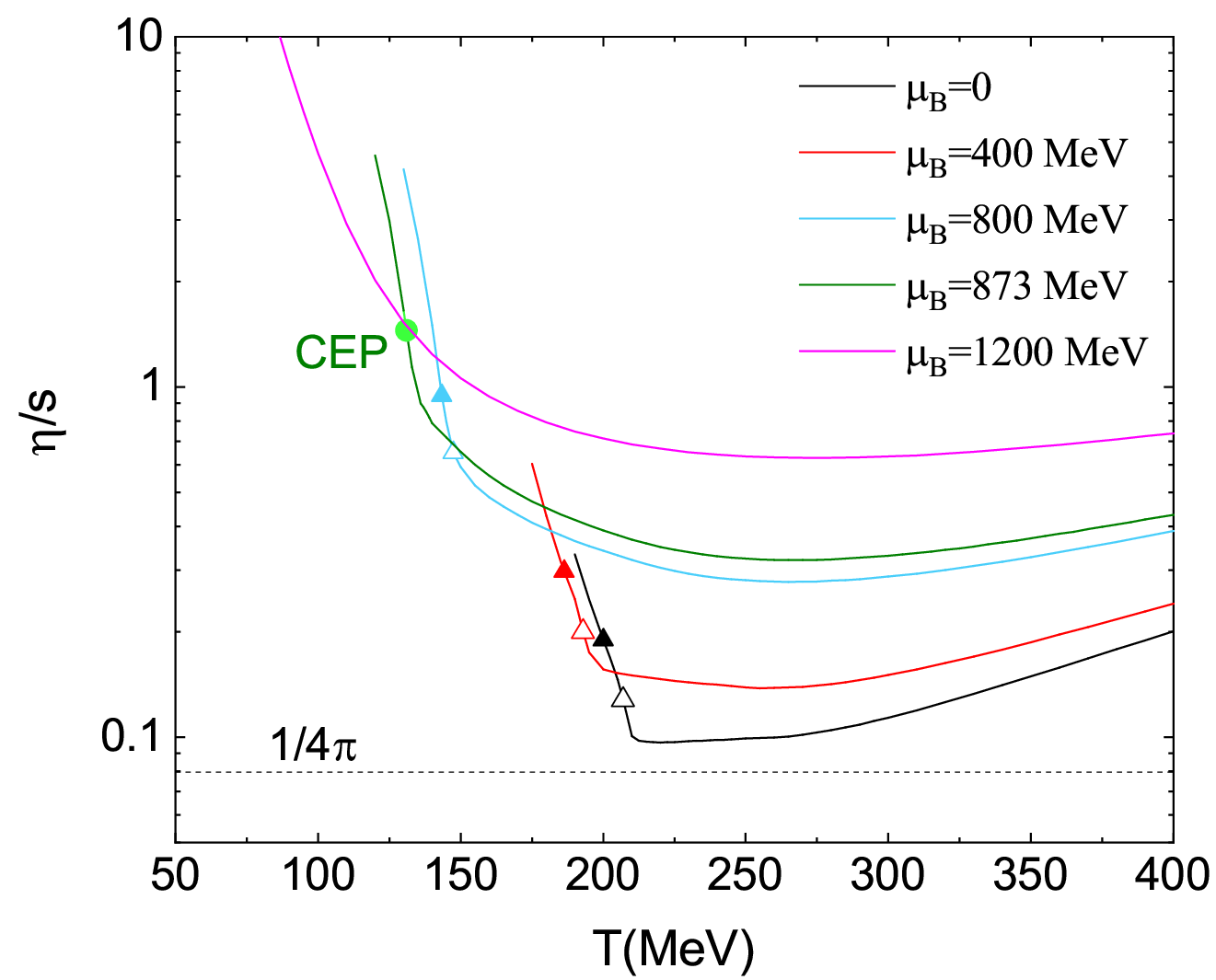}
	\caption{Shear visicosity to entropy density ratio $\eta/s$  as functions of temperature at different $\mu_B$ in the PNJL model. The dot, solid triangles and hollow triangles represent  the $\eta/s$ at the points where these $\mu_B(T)$ lines intersect the CEP, chiral crossover phase transition line and pion Mott transition line, respectively.}
	\label{fig:6} 
\end{figure}

In the following, we elaborate on the shear viscosity of QGP under different conditions. First of all, we present in Fig.~\ref{fig:6} the shear viscosity to entropy density ratio as functions of temperature at different chemical potentials.  The dot, solid triangles and hollow triangles represent respectively the $\eta/s$ at the points where these $T(\mu_B)$ lines intersect the CEP, chiral crossover phase transition line and pion Mott transition line.
For $\mu_B$=0, the value of $\eta/s$  are very small when the temperature is close to and above the meson Mott phase transition.
The minimum of $\eta/s$ is about 0.1, close to the KSS limit, which indicates that the QGP medium is a nearly ``perfect" fluid in the very high energy collisions. The value of $\eta/s$ increases with the rising chemical potential. It can be observed from Fig.~\ref{fig:6} that $\eta/s$ is larger than 1 in the critical region, and the value will continually increase at lower temperature and larger chemical potential. 

The minimum of  $\eta/s$ at $\mu_B=0$ is  closer to the Mott phase transition~(meson dissociation) than the chiral phase transition of $u,d$ quark, and  $\eta/s$ increases rapidly with the occurrence of hadronization. In Fig.~\ref{fig:6} some values of  $\eta/s$  below the chiral phase line are also preserved to display the variation of $\eta/s$ near the chiral crossover phase transition region. This figure clearly indicates that the properties of $\eta/s$ much depends on temperature, in particular near the phase transition.
However we should keep in mind that the collisions between hadrons dominate the transport process in the hadronic phase at low temperature, which is not considered in the present work.  The QGP phase in this study means
the deconfined QCD medium within the PNJL model.
It can be observed from  Fig.~\ref{fig:6} that the location of the minimum  $\eta/s$ at finite chemical potential become more and more flat, and the relationship with phase transitions is not as distinct as that of $\mu_B=0$. 

\begin{figure}[htbp]
	\includegraphics[scale=0.37]{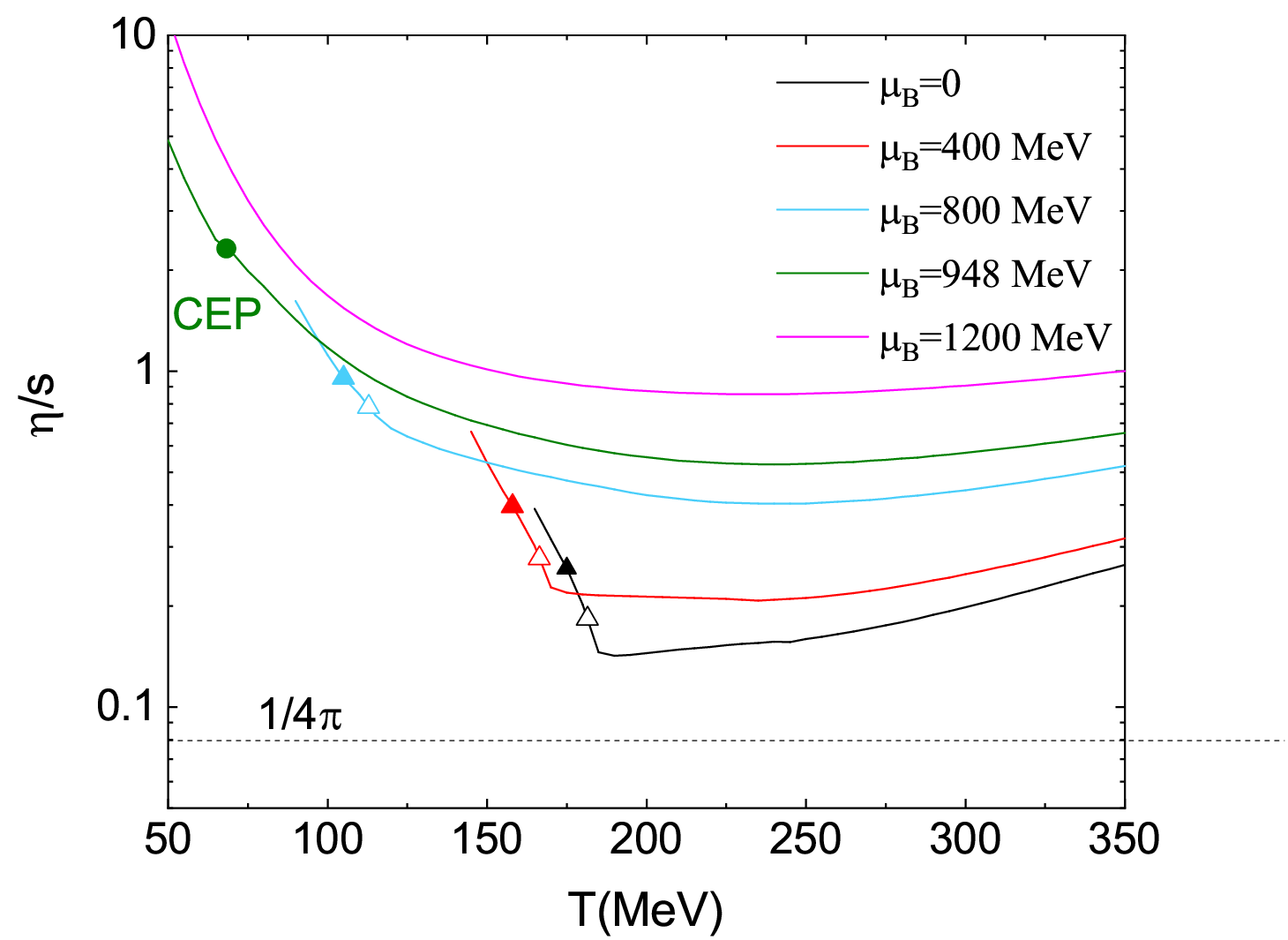}
	\caption{Shear visicosity to  entropy density ratio $\eta/s$  as functions of temperature at different $\mu_B$ in NJL model. 
The dot, solid triangles and hollow triangles represent the $\eta/s$ at the points where these $T(\mu_B)$ lines intersect the CEP, chiral crossover phase transition line and pion Mott transition line, respectively.	}
	\label{fig:7} 
\end{figure}

It is interesting to compare the PNJL results against those in the NJL model with a lower $T_C$. We present in Fig.~\ref{fig:7}  the temperature-dependent $\eta/s$  for a series of baryon chemical
potential $\mu_{B}=0$, $400$, $800$, $948$ and $1200$ MeV in the NJL model.  Compare Fig.~\ref{fig:6} and Fig.~\ref{fig:7}, we can see that the two models give similar behavior of  $\eta/s$. For small chemical potentials, the $\eta/s$ in  the NJL model also exhibit a characteristic non-monotonic behavior with rising temperature, and the minimum appears at a temperature higher than the chiral crossover phase transition temperature~(marked with the solid triangle). For the case of vanishing baryon chemical potential the minimum value of $\eta/s$ in the NJL model is slightly larger than that in the PNJL model, but are still close to 0.1. At the CEP, the values of $\eta/s$ are above 1 in both models, which indicates to some extent of the universality of shear viscosity near the critical region of strongly interacting matter.

\begin{figure}[htbp]
	\centering
	\includegraphics[scale=0.37]{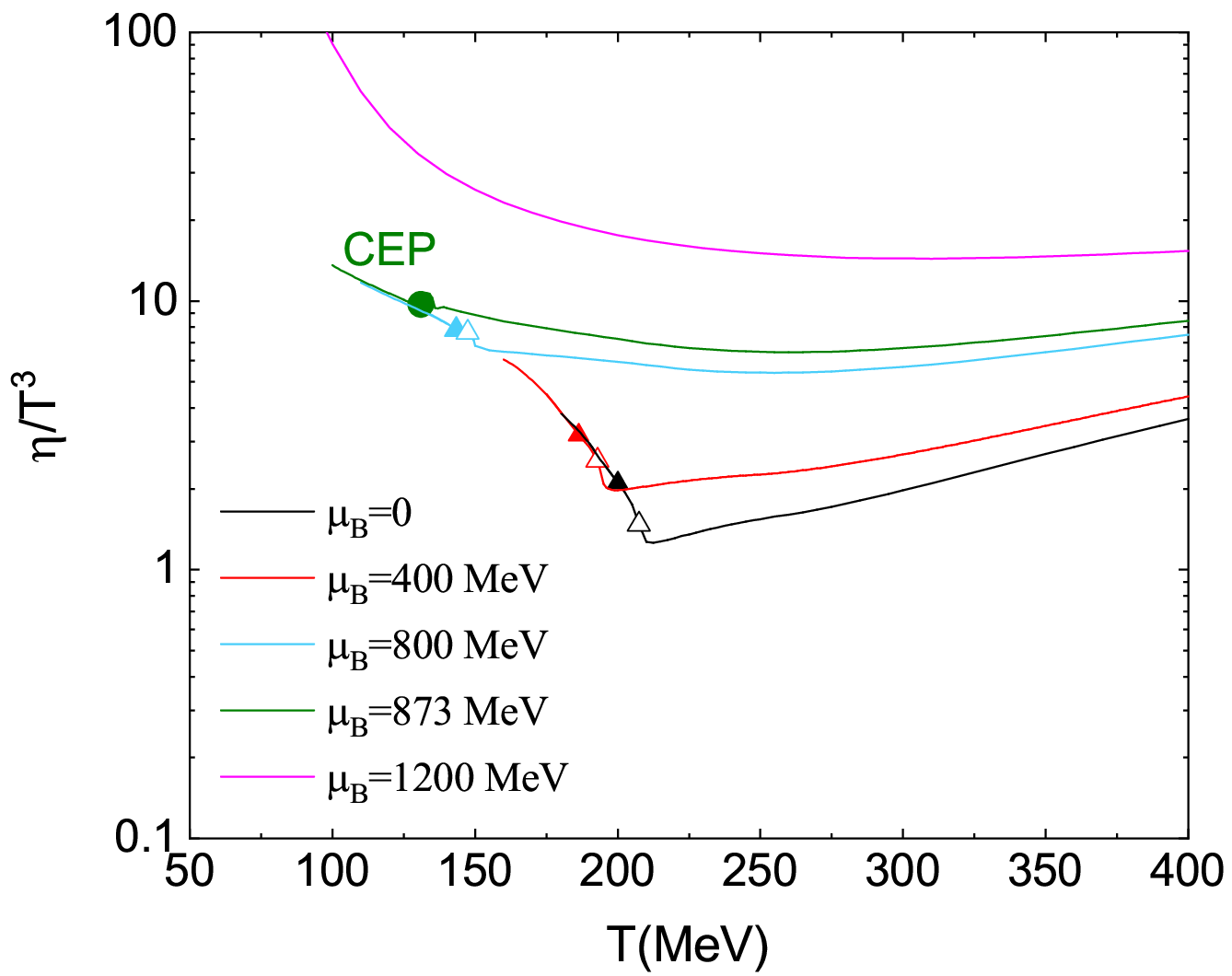}
	\caption{The scaled shear viscosity $\eta/T^{3}$ as functions of temperature at different chemical potential $\mu_{B}=0$, $400$, $800$, $873$ and $1200\,$MeV in the PNJL model. The dot, solid triangles and hollow triangles represent  the $\eta/T^3$ at the points where these  $\mu_B(T)$ lines intersect the CEP, chiral crossover phase transition line and pion Mott transition line, respectively.}
	\label{fig:8} 
\end{figure}

\begin{figure}[htbp]
	\centering
	\includegraphics[scale=0.37]{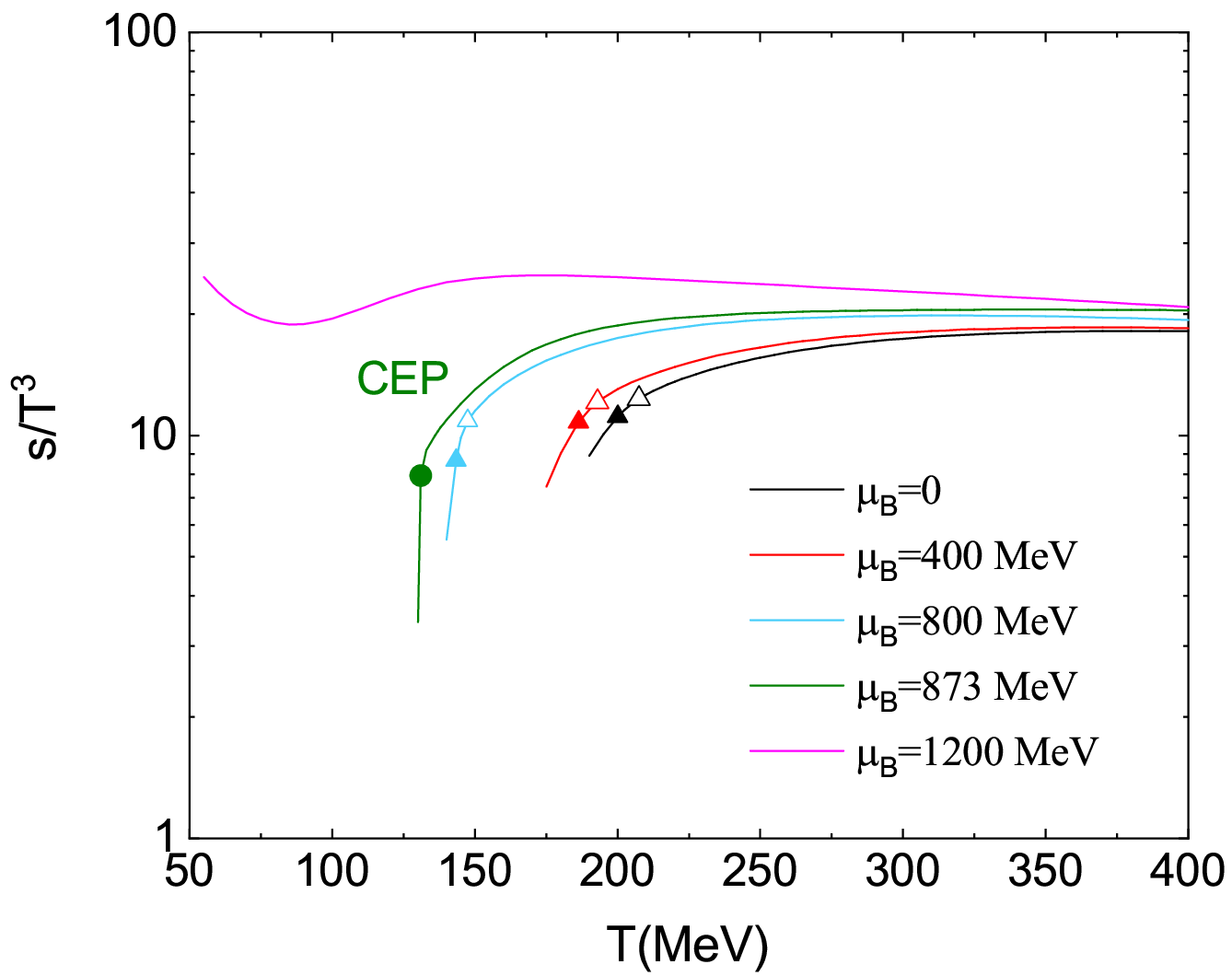}
	\caption{The scaled entropy density $s/T^{3}$ as function of temperature at different $\mu_{B}$ in the PNJL model.The dot, solid triangles and hollow triangles represent the $s/T^3$ at the points where these  $\mu_B(T)$ lines intersect the CEP, chiral crossover phase transition line and pion Mott transition line, respectively.}
	\label{fig:9} 
\end{figure}

To understand the behavior of $\eta/s$ in the $T-\mu_B$ plane in Fig.~\ref{fig:6} derived in the PNJL model, we  decompose the $\eta/s$ into shear viscosity coefficient $\eta$ and entropy density $s$. The scaled shear viscosity $\eta/T^{3}$ and entropy density $s/T^{3}$ are plotted in Figures~\ref{fig:8} and~\ref{fig:9}.
We can see that the $\eta/T^{3}$ for a given small chemical potential has a minimum with the increasing temperature and the corresponding  $s/T^{3}$ only changes slightly at temperature higher than the Mott phase transition. Therefore, the existence of a minimum of $\eta/s$ is mainly attributed to the temperature-dependent behavior of $\eta/T^{3}$. Besides, below the chiral phase temperature, the value of $\eta/T^{3}$ increases quickly with the decrease of temperature, and meanwhile the value of $s/T^{3}$ declines sharply. They together lead to the quick enhancement of $\eta/s$ below the chiral phase transition temperature, as shown in Fig.~\ref{fig:6}.


\begin{figure}[htbp]
	\begin{center}
		\includegraphics[scale=0.37]{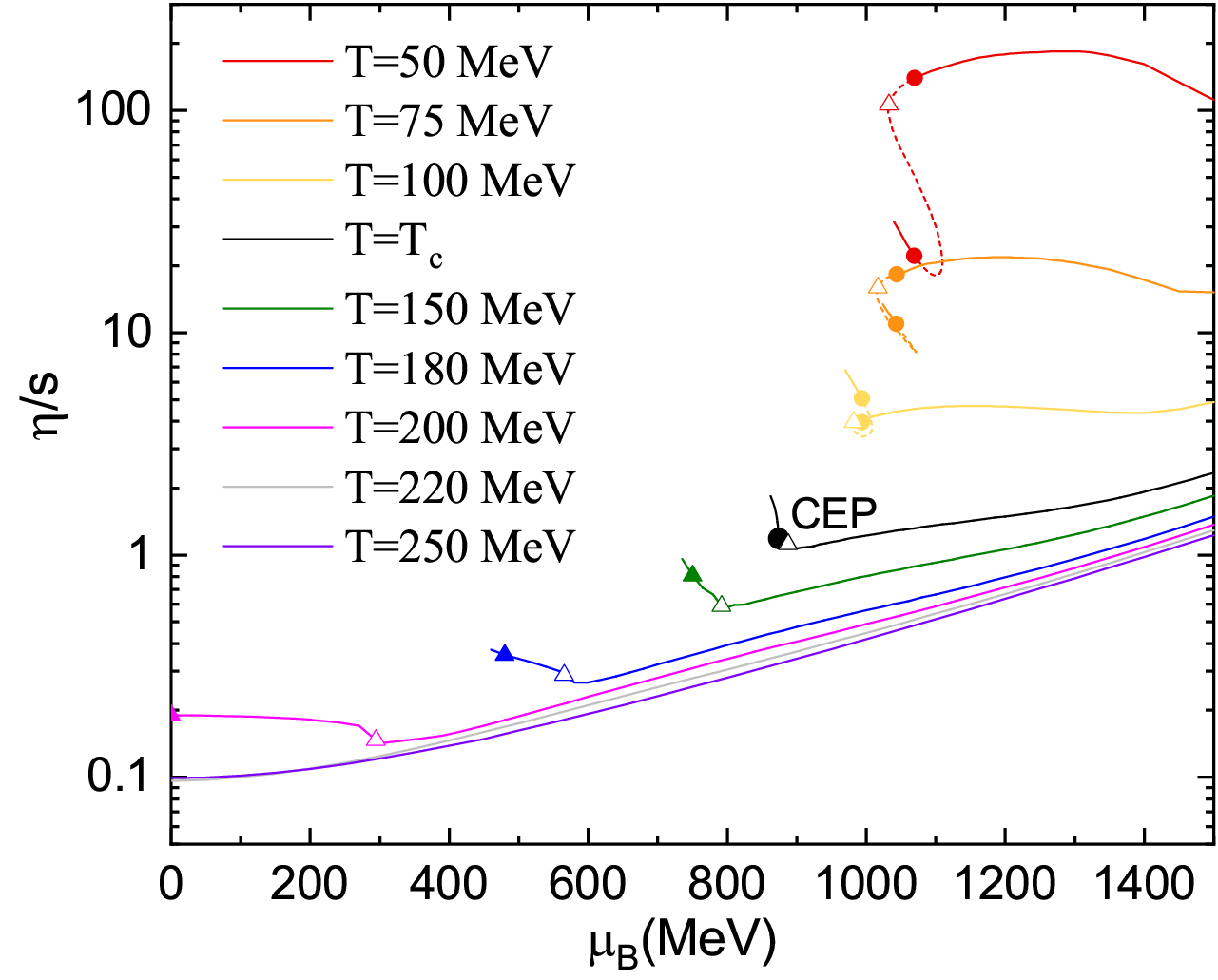}
		\caption{\label{fig:10}  $\eta/s$  as functions of baryon chemical potential at different temperatures  in the PNJL model. The solid triangles,  dots, and hollow triangles represent  the $\eta/s$ at the points where these $T(\mu_B)$ lines intersect the chiral corssover transition line, the first-order chiral phase transition line and pion Mott transition line, respectively. For $T=50, 75, 100\,$MeV, the two dots on each isotherm correspond to respectively the high-density chiral restoration phase and low-density chiral breaking phase on the boundaries (binodal line) of the first-order phase transition. 
		}
	\end{center}
\end{figure}

Fig.~\ref{fig:10} describes the behaviors of $\eta/s$ as functions of chemical potential at several temperatures lower than or equal to $250\,$MeV. For the temperature higher than that of the chiral phase transition of $u(d)$ quark, e.g.,  $T=220$ or $250\,$MeV, it can be seen  that $\eta/s$ monotonically increases with the rising chemical potential.  The nonmonotonicity appears  with the decrease of temperature, and  the location of the minimum of $\eta/s$  gradually approaches to the chiral crossover phase transition (solid triangle in Fig.~\ref{fig:10}), particularly in the critical region. 

Fig.~\ref{fig:10} also presents the values of $\eta/s$ at $T=100,75,50$ MeV where the first-order phase transition associated with a spinodal structure is involved. The solid dots on each isotherm with the same color indicates the value of $\eta/s$ on the boundaries of the first-order phase transition. From higher chemical potentials to lower ones, the first solid dot encountered on each curve corresponds to the high-density phase of the first-order chiral phase transition (in chiral restoration phase), and the second solid dot corresponds to the low-density phase of the chiral phase transition (in chiral breaking phase). This figure indicates that the values of $\eta/s$ are quite different in the  chiral broken and chiral restored phases at lower temperature. The interval between the two dots on each isotherm  corresponds to the metastable and unstable region. 

Different from the first-order transition, the hollow triangles in Fig.~\ref{fig:10} show that each $T(\mu_B)$ line for  $T=100,75,50\,$ MeV crosses only once the pion Mott phase transition line in the metastable or unstable phase. This can be understood in conjunction with Fig.~\ref{fig:3}. As illustrated in Fig.~\ref{fig:3}, the pion Mott transition line intersects solely with the high-density boundary of the first-order transition and then enters the metastable and unstable phases as temperature decreases. Since each point on the Mott line is defined by a specific $(T, \rho_B)$ value which corresponds to a unique set of  $(T, \mu_B)$.  Therefore the  $\eta/s$ curve as a function of baryon chemical potential at a fixed temperature~(e.g., $T=100\,$MeV) in Fig.~\ref{fig:10} intersects the Mott transition line at only one point.

\begin{figure}[htbp]
	\begin{center}
		\includegraphics[scale=0.37]{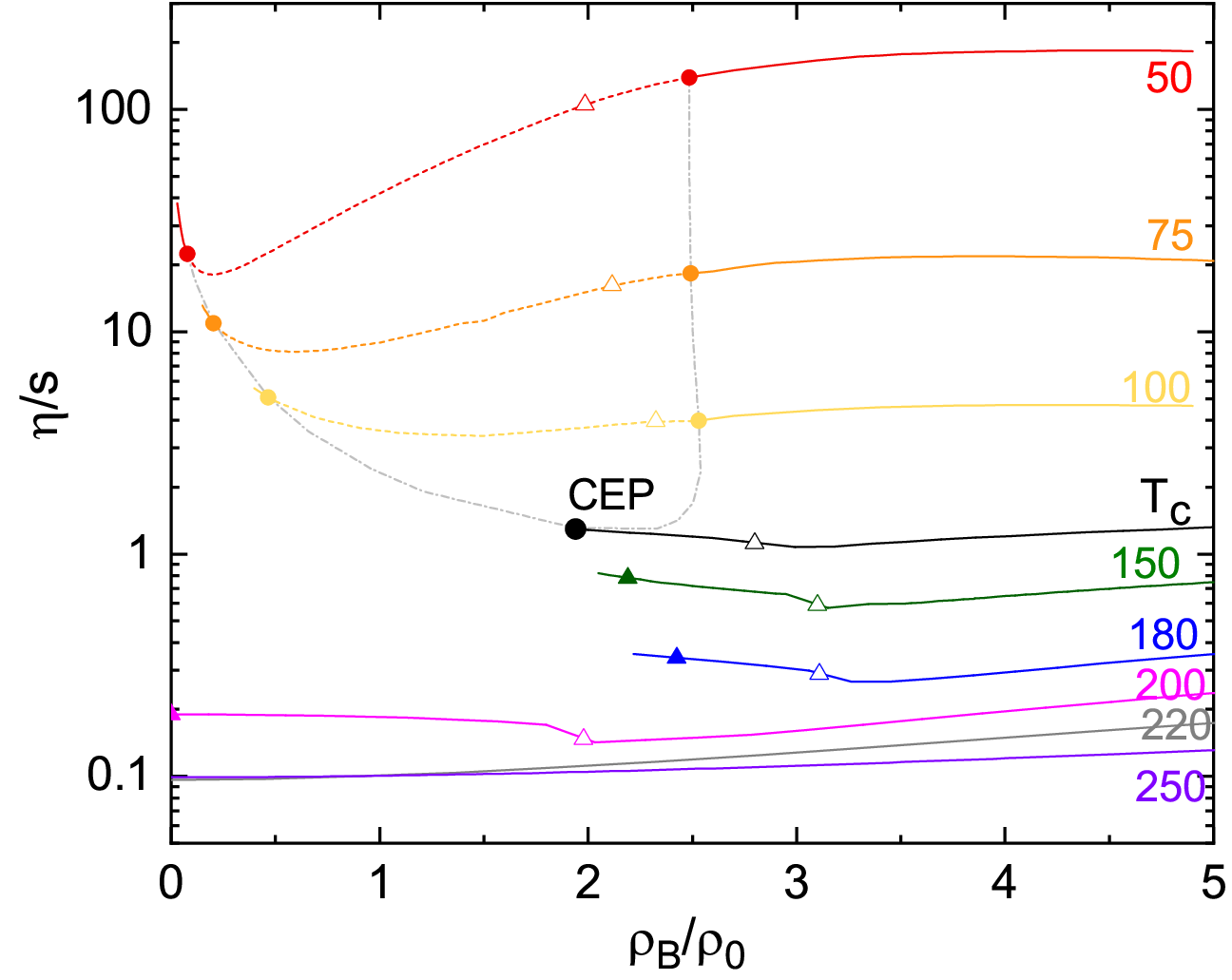}
		\caption{\label{fig:11}$\eta/s$  as functions of net baryon density  at different temperatures in the PNJL model. The solid triangles,  dots, and hollow triangles represent   the $\eta/s$ at the points where these $T(\rho_B)$ lines intersect the chiral corssover transition line, the first-order chiral phase transition line and pion Mott transition line, respectively. For $T=50, 75, 100\,$MeV, the two dots on each isotherm correspond to respectively the high-density chiral restoration phase and low-density chiral breaking phase on the  binodal line of the first-order phase transition. }
	\end{center}
\end{figure}

Corresponding to Fig.~\ref{fig:10}, Fig.~\ref{fig:11} presents the value of $\eta/s$ in the $T-\rho_B$ plane for different temperatures.  The dashed curve connecting different isotherms is the  boundary of the first-order phase transition. The behaviors of $\eta/s$ at $T=200$ MeV or higher temperatures  are similar to those in the $T-\mu_B$ plane plotted in Fig.~\ref{fig:10}.
However, compared to Fig.~\ref{fig:10}, in the  $T-\rho_B$ plane the distance between the net baryon density of the minimum  $\eta/s$ and that of the chiral crossover phase transition (marked with the solid triangle) appears to be enlarged when approaching the critical region of the CEP from the high-temperature side. The reason is that the net baryon density is sensitive to chemical potential in the critical region. Small variation of chemical potential will lead to a big change of net baryon density. 
For the first-order phase transition, the difference of $\eta/s$ in the two phases is small near the CEP, since the dynaical quark masses of quarks are slight different.  However, it can be seen that at lower temperature, e.g., $T=50$ MeV, $\eta/s$ in the chiral restored QGP phase is much larger than the chiral breaking phase.
It also indicates  the shear viscosity to entropy density ratio near the first-order phase transition at lower temperatures very depends on the net baryon density.

\begin{figure}[htbp]
	\begin{center}
		\includegraphics[scale=0.35]{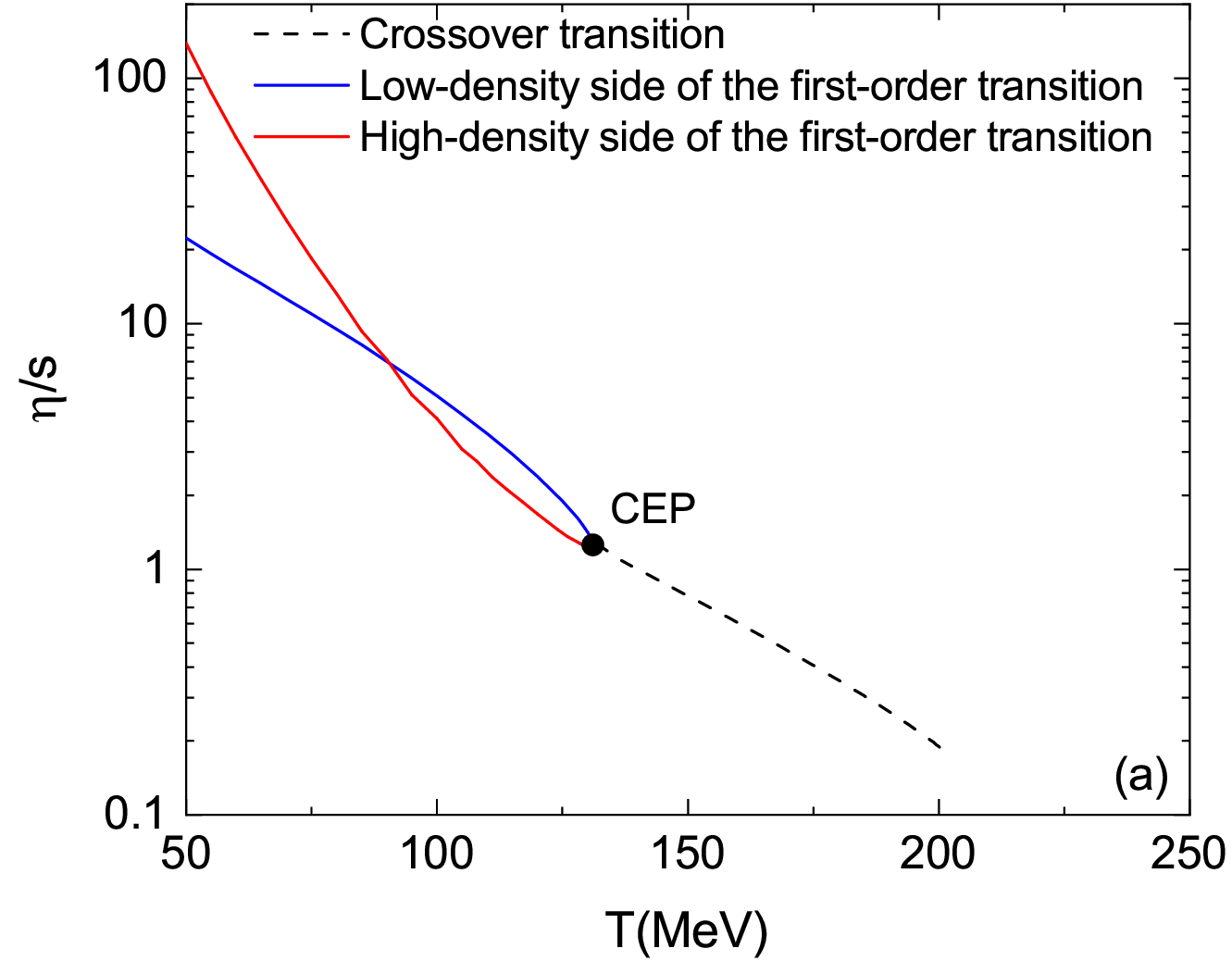}
		\includegraphics[scale=0.35]{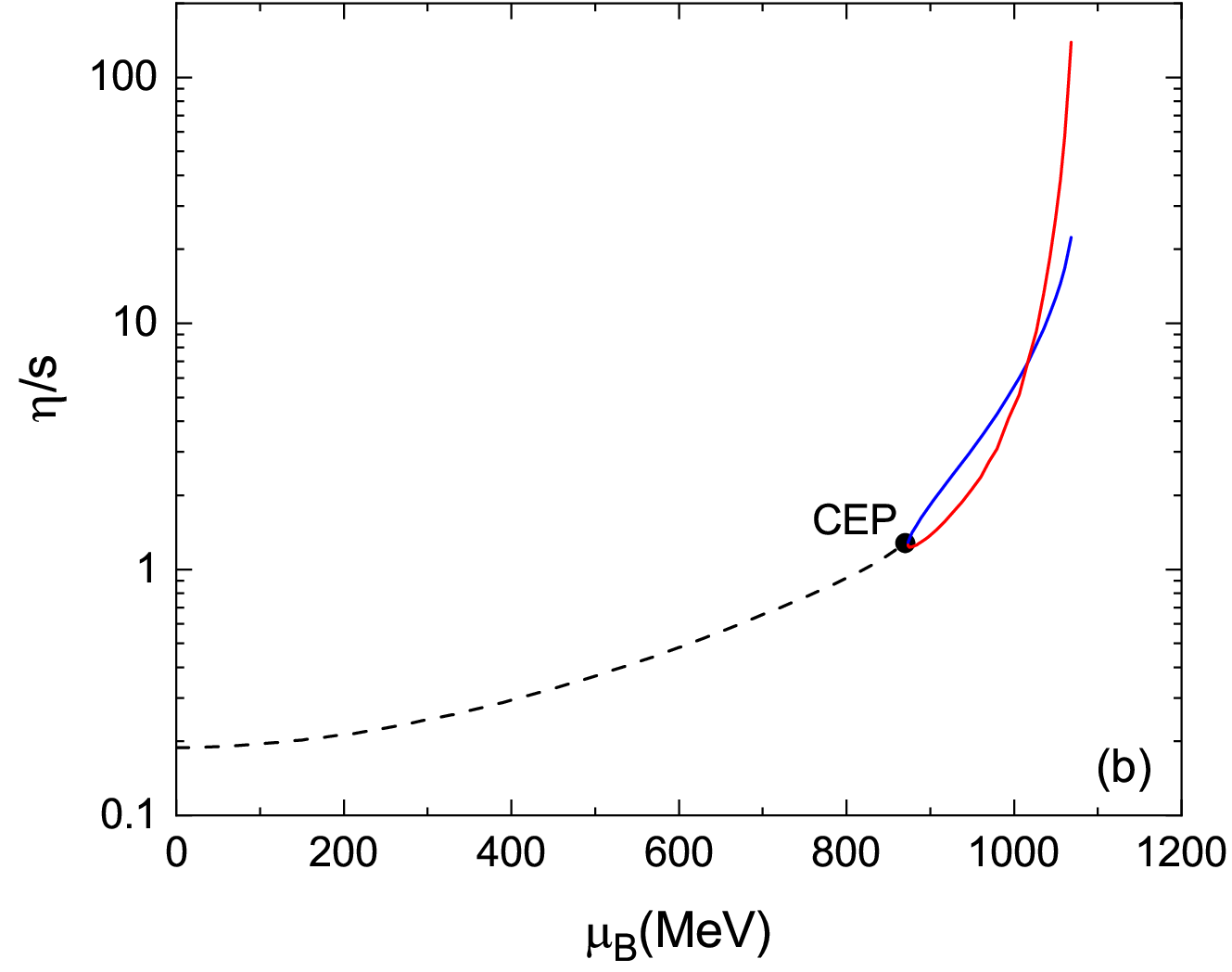}
		\includegraphics[scale=0.35]{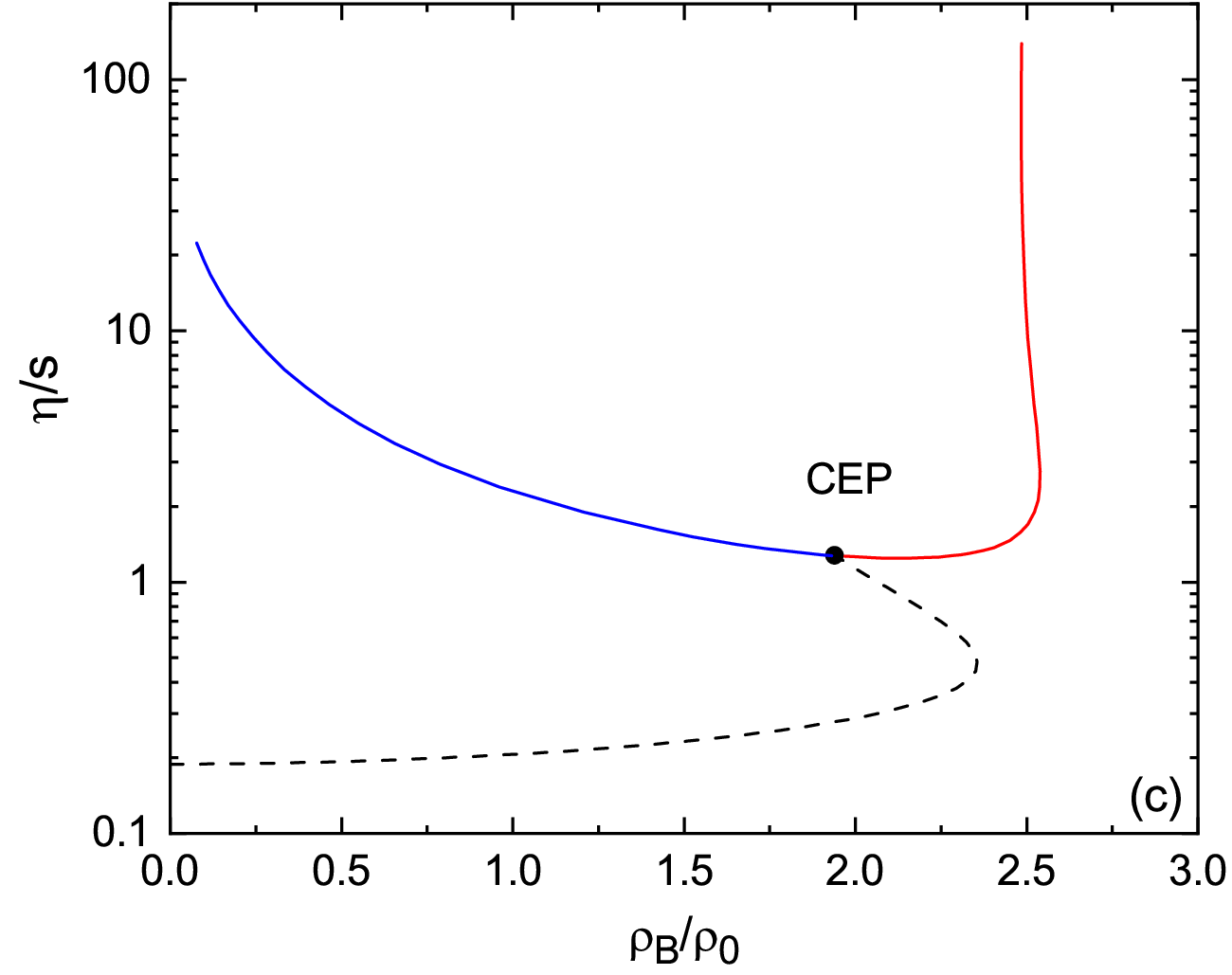}
		\caption{\label{fig:12} The value of $\eta/s$  on the phase boundaries including the chiral corssover and first-order phase transitions  in the PNJL model.
		}
	\end{center}
\end{figure}

Fig.~\ref{fig:10} and Fig.~\ref{fig:11} also present that, in high-density region, the value of $\eta/s$ increases with the decline of temperature, which is qualitatively consistent with the variation of shear viscosity of common liquids. Importantly, the change of $\eta/s$ with the physical conditions inevitably causes us to pay more attention to the viscosity of QGP medium at finite temperature and density. The value of $\eta/s$ is crucial in diagnosing the QCD phase transition signals in the hydrodynamical simulation of heavy-ion collisions with colliding energies at several GeV, close to the lower side of the beam energy scan at RHIC STAR.  that $\eta / s \approx 0.65 \pm 0.15$ in Au+Au reactions at $\sqrt{s}=2.4$ GeV~\cite{reichert2021first}.

We plot in Fig.~\ref{fig:12} the value of $\eta/s$ on the boundaries of the chiral phase transition of $u, d$ quarks.  Fig.~\ref{fig:12}(a), (b) and (c) present the behavior of $\eta/s$ as functions of temperature, baryon chemical potential and net baryon density, respectively. Since the relation between $\rho_B$ and $\mu_B$ on the crossover phase transition line near the critical region is nonmonotonic as indicated in Fig.~\ref{fig:2} and Fig.~\ref{fig:3}, Fig.~\ref{fig:12}(c) shows a similar structure of $\eta/s$.  For the first-order phase transition, Fig.~\ref{fig:12}(a) also indicates that $\eta/s$ in the chiral restored QGP phase is smaller than that in the chiral breaking  phase in  the region not far away from the critical region. This situation reverses as the temperature decreases~(correspondingly, the chemical potential increases as shown in Fig.~\ref{fig:12}(b)). Such a feature reflects that the value of $\eta/s$ in the two phase of the chiral first-order phase transition are affected by temperature, chemical potential~(density) and QCD phase transition together.

\begin{figure}[htbp]
	\begin{center}
		\includegraphics[scale=0.37]{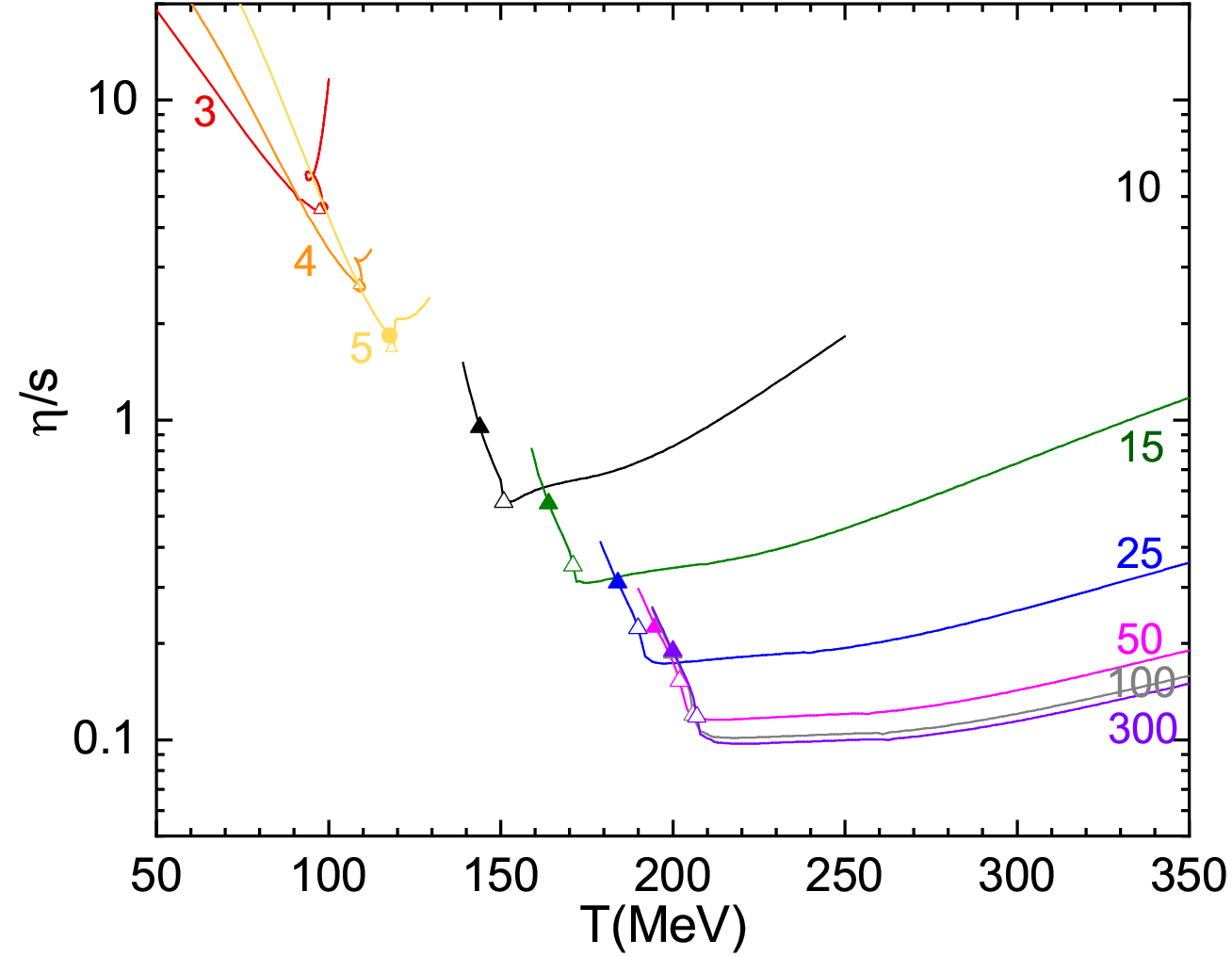}
		\caption{\label{fig:13} $\eta/s$  on different isentropic trajectories  in the PNJL model. The isentropic trajectories are plotted in Fig.~\ref{fig:1}. 
			The dot, solid triangles and hollow triangles represent the $\eta/s$ at the points where these isentropic lines intersect the high-density side of the first-order phase transition line, chiral crossover phase transition line and pion Mott transition line, respectively.}
	\end{center}
\end{figure}

We demonstrate in Fig.~\ref{fig:13} the behavior of $\eta/s$  along the isentropic trajectories as plotted in Fig.~\ref{fig:1} and Fig.~\ref{fig:2}.  The dot, solid triangles and hollow triangles represent the $\eta/s$ at the points where these isentropic line intersect the high-density side of the first-order phase transition line, chiral crossover phase transition line and pion Mott transition line, respectively.
It is helpful to understand the result in conjunction with QCD phase diagrams shown in  Fig.~\ref{fig:1} and Fig.~\ref{fig:2}.
It can be seen in Fig.~\ref{fig:13} that $\eta/s$ has a minimum on each isentropic trajectory. 
For the case of larger $s/\rho_B$, the isentropic lines, as shown in Fig.~\ref{fig:1}, are almost parallel to the temperature axis. Thus, it is straightforward to see that the behavior of the corresponding shear viscosity with growing temperature is similar to that with a fixed chemical potential (in the region of small chemical potential). 
For the case of smaller $s/\rho_B$, 
the isentropic curves, as shown in  Fig.~\ref{fig:2}, traverse the first-order phase transition from low to high density, including the low-density chiral symmetry broken phase and the high-density chiral symmetry restored phase (with metastable and unstable phases existing between them). Due to the spinodal structure inherent in the first-order phase transition, the isentropic curve exhibits a complex trajectory in the $T-\mu_B$ phase diagram, as illustrated in  Fig.~\ref{fig:1}.
For $s/\rho_B=3,4,5$, the value of $\eta/s$ changes rapidly near the first-order phase transition with the associated spinodal structure.
The numerical results show that the minimum value of $\eta/s$ on each curve occurs near the Mott transition, which is close to the high-density boundary of the first-order phase transition, as indicated by the solid dots in Fig.~\ref{fig:10}.

\subsection{Electric conductivity of quark matter}

The electrical conductivity ($\sigma$) is a key parameter characterizing the charge transport properties of QGP. It directly influences the electromagnetic response and evolution behavior of the plasma, and is related to  physical issues such as phase transition dynamics and electromagnetic field evolution. In this section, we will explore the electrical conductivity of quark matter at finite temperature and density, in particular its behavior near the Mott and chiral phase transition with the inherent spinodal structure.

\begin{figure}[htbp]%sigma/T (T) fixed mu.
	\centering
	\includegraphics[scale=0.37]{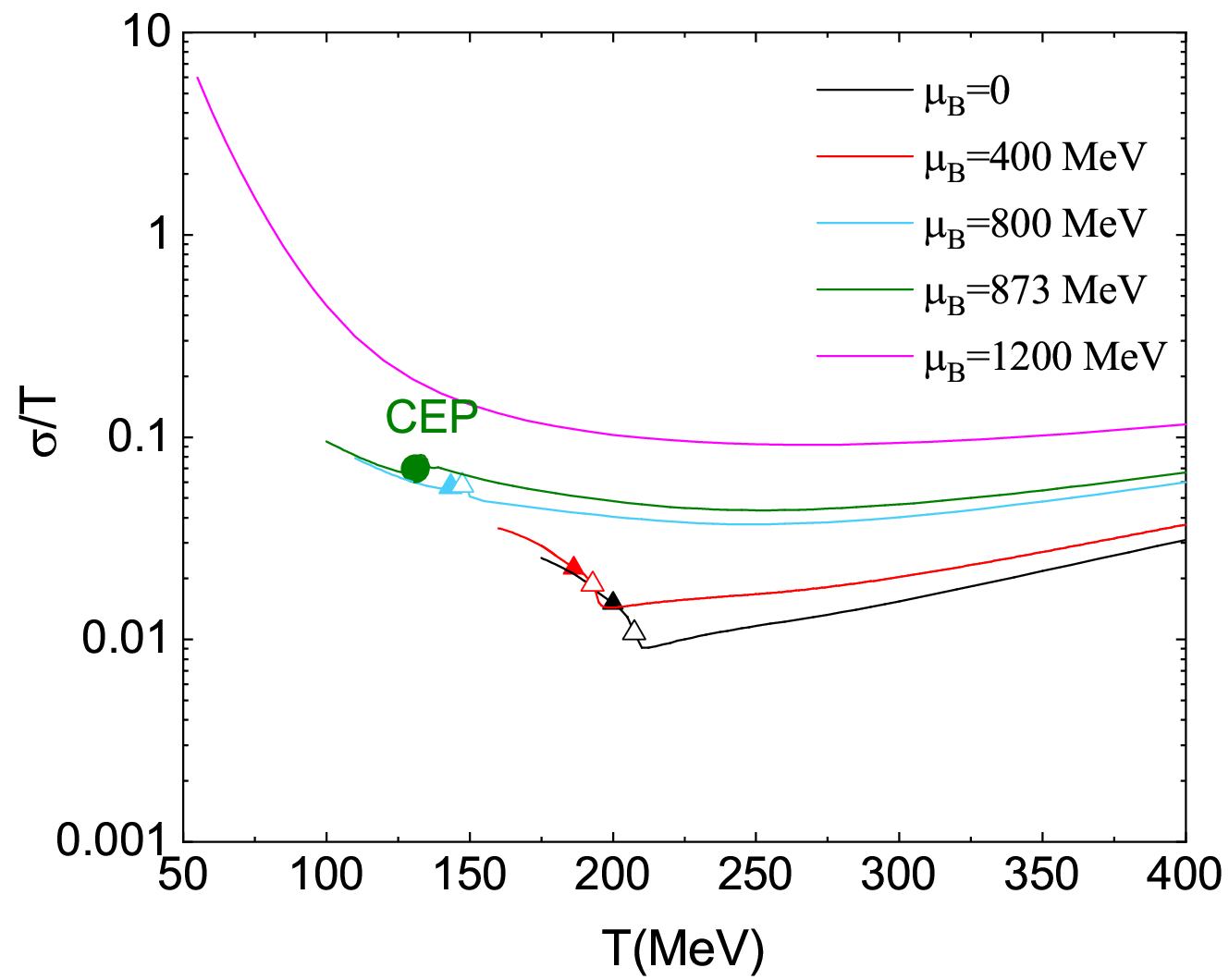}
	\caption{Electric conductivity to temperature ratio $\sigma/T$ as function of temperature at different chemical potential $\mu_{B}=0$, $400$, $800$, $873$ and $1200\,$ MeV in the PNJL model. The dot, solid triangles and hollow triangles represent the $\sigma/T$ at the points where these  $\mu_B(T)$ lines intersect the CEP, chiral crossover phase transition line and pion Mott transition line, respectively.}
	\label{fig:14} 
\end{figure}

We first display in Fig.~\ref{fig:14} the ratio of the electrical conductivity to temperature, $\sigma/T$, as a function of temperature for different baryon chemical potentials. At small chemical potentials, a distinct minimum in $\sigma/T$ appears near the Mott transition, beyond which $\sigma/T$ is observed to increase with temperature. This trend is consistent with the lattice QCD calculations~\cite{Aarts2020dda, Amato2013naa, Aarts2014nba, Brandt2015aqk}. Along with the chiral crossover phase transition, the value of $\sigma/T$ rises with decreasing temperature. This behavior aligns with hadronic model calculationss~\cite{Cassing2013iz}.
The characteristics of the electrical conductivity resemble those of the shear viscosity near the phase transition region. Specifically, both $\eta/s$ and $\sigma/T$ exhibit a concave-down structure with a pronounced minimum near the Mott transition, indicating that intense collisions and resonances hinder the transport of momentum and electric charge.
At large chemical potentials, a significant enhancement of $\sigma/T$ is observed, which is primarily attributed to the increase in quark number density.

\begin{figure}[htbp]
	\begin{center}
		\includegraphics[scale=0.37]{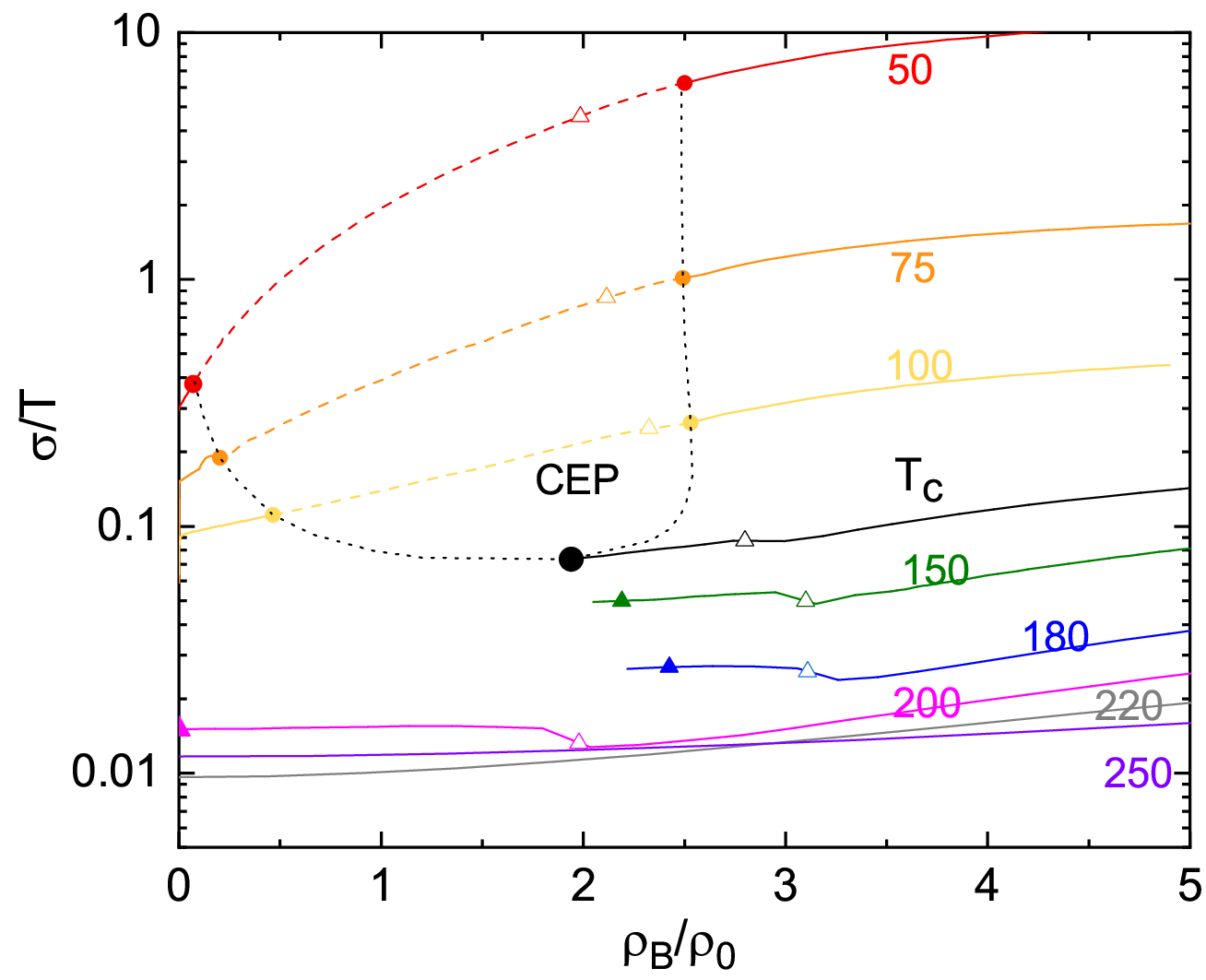}
		\caption{\label{fig:15}$\sigma/T$  as functions of net baryon density  at different temperatures in the PNJL model. The solid triangles,  dots, and hollow triangles represent  the $\sigma/T$ at the points where these $T(\rho_B)$ lines intersect the chiral corssover transition line, the first-order chiral phase transition line and pion Mott transition line, respectively. For $T=50, 75, 100\,$MeV, the two dots on each isotherm correspond to respectively the high-density chiral restoration phase and low-density chiral breaking phase on the  binodal line of the first-order phase transition. }
	\end{center}
\end{figure}

In Fig.~\ref{fig:15}, we present the behavior of $\sigma/T$ as a function of net baryon  density at different temperatures. Above the critical temperature, the temperature dependence of $\sigma/T$ is similar to that of $\eta/s$. Below the critical temperature, $\sigma/T$ gradually increases from low to high density. The two solid dots on each isotherm mark the boundaries of the first-order phase transition: the dot at lower density corresponds to the chiral symmetry broken phase, while the dot at higher density corresponds to the chiral symmetry restored phase. Evidently, for the first-order phase transition, the value of $\sigma/T$ in the chiral symmetry restored phase is lower than that in the chiral symmetry broken phase.

\begin{figure}[htbp]
	\begin{center}
		\includegraphics[scale=0.37]{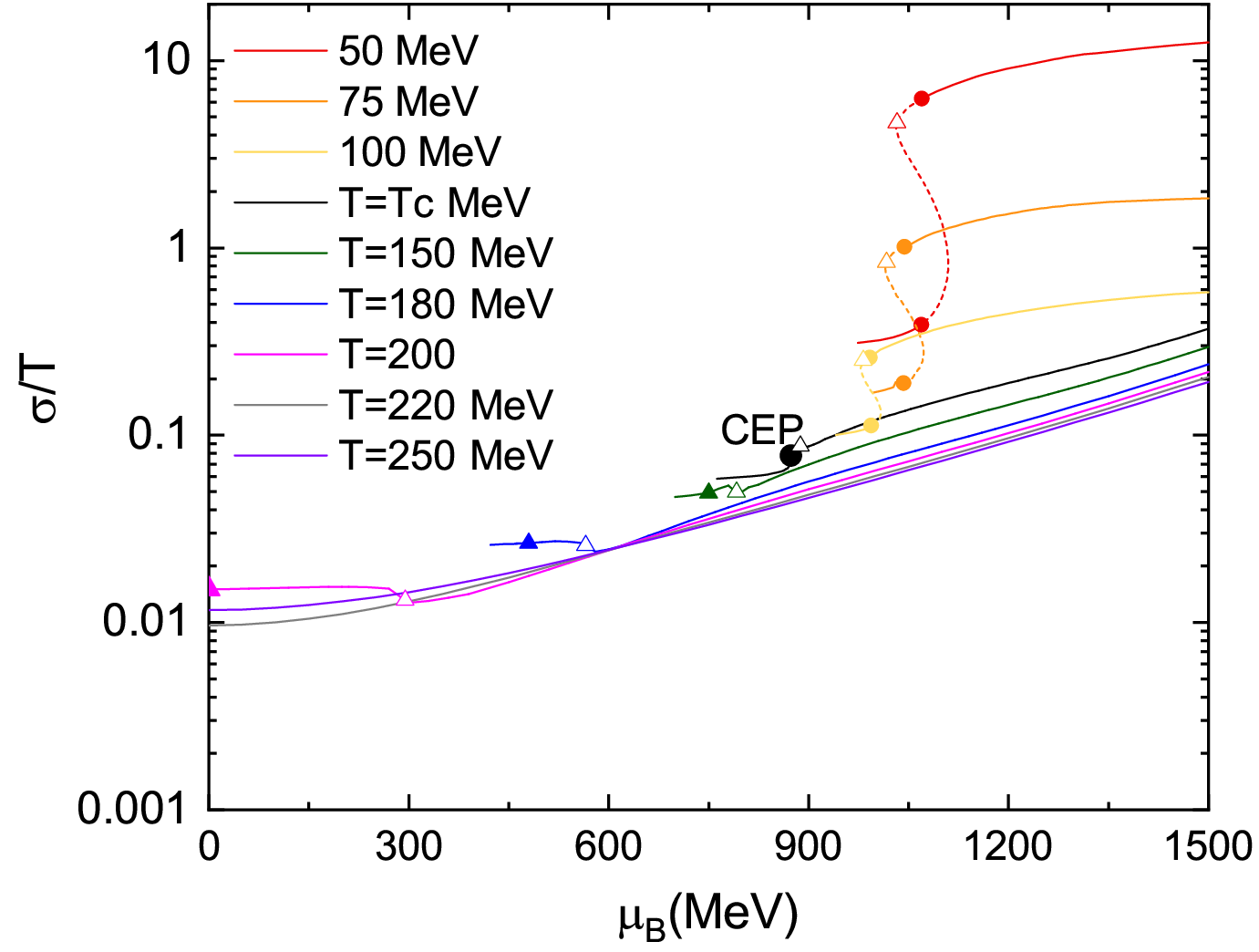}
		\caption{\label{fig:16}$\sigma/T$  as functions of baryon chemical potential at different temperatures  in the PNJL model. The solid triangles,  dots, and hollow triangles represent  the $\sigma/T$ at the points where these $T(\mu_B)$ lines intersect the chiral corssover transition line, the first-order chiral phase transition line and pion Mott transition line respectively. For $T=50, 75, 100\,$MeV, the two dots on each curve correspond to respectively the high-density chiral restoration phase and low-density chiral breaking phase on the  binodal line of the first-order phase transition. }
	\end{center}
\end{figure}

We correlate the behavior of $\sigma/T$ with net baryon number density by projecting it onto the $T$-$\mu_B$ phase diagram in Fig.~\ref{fig:16}. The dashed curves on the isotherms delineate the metastable and unstable regions associated with the first-order phase transition. Fig.~\ref{fig:15} and Fig.~\ref{fig:16} reveal a significant enhancement of the electrical conductivity in the low-temperature, high-density regime (large $\mu_B$) relative to the high-temperature, low-density region.

\begin{figure}[htbp]
%	\begin{center}
		\includegraphics[scale=0.35]{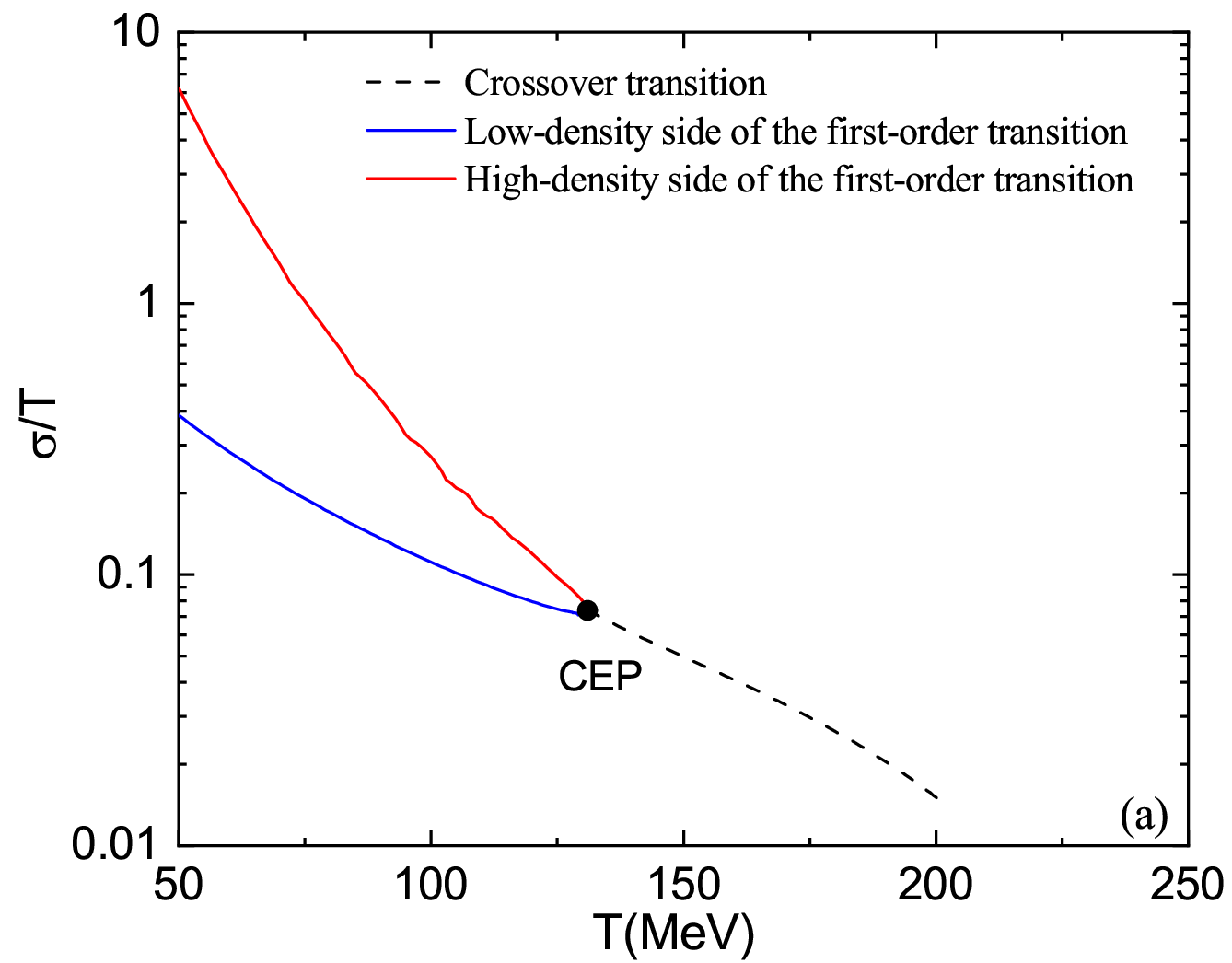}
		\includegraphics[scale=0.35]{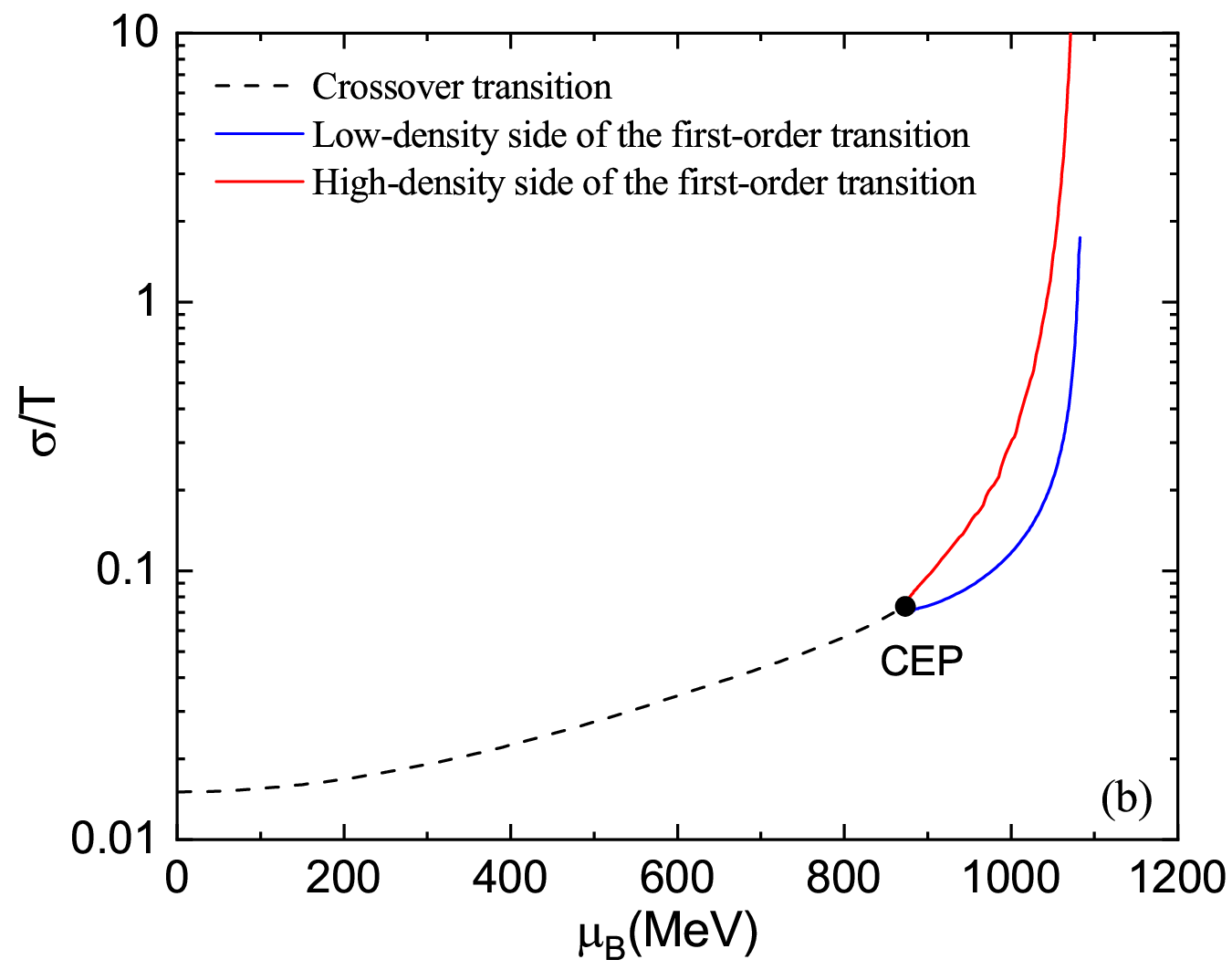}
		\includegraphics[scale=0.35]{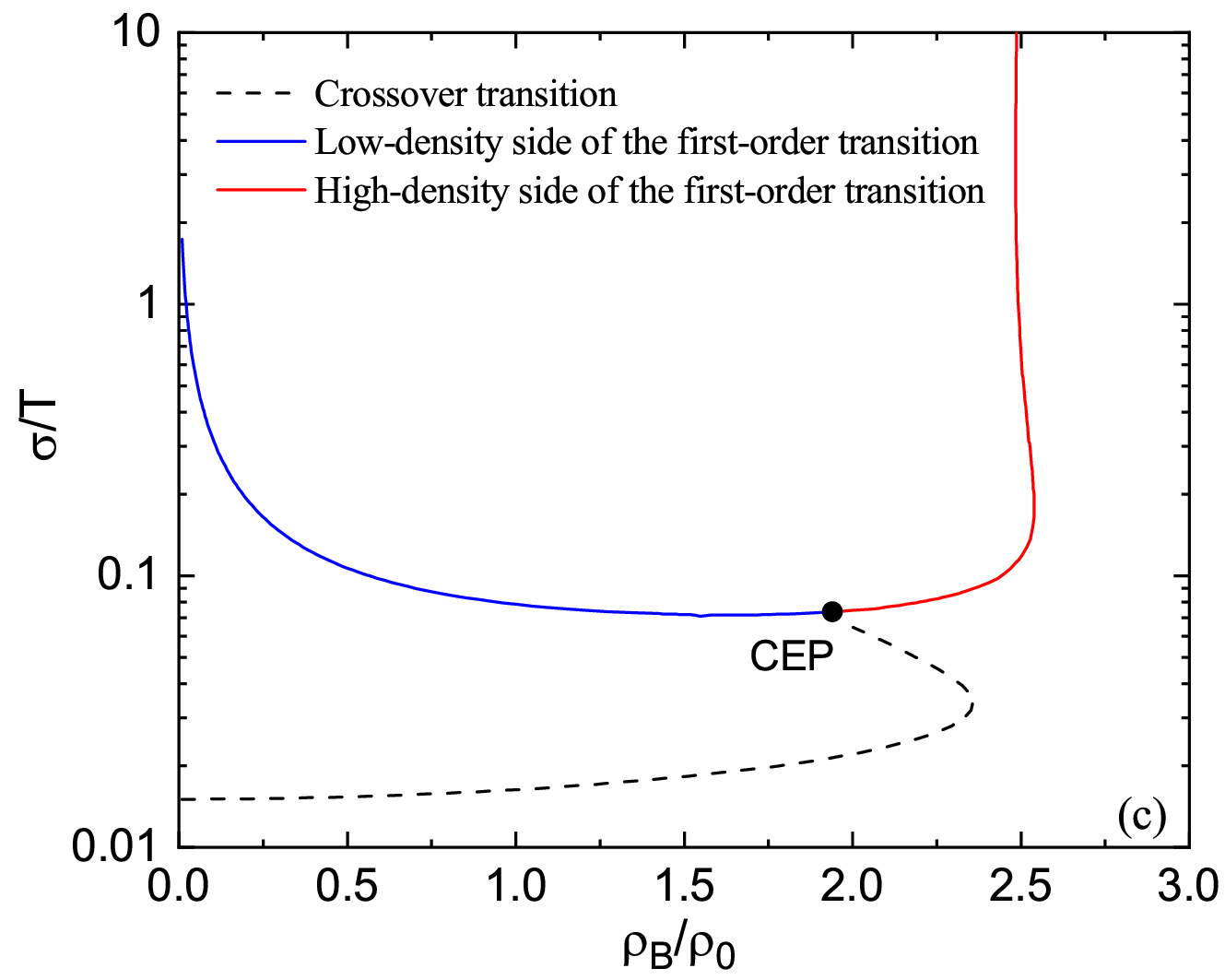}
		\caption{\label{fig:17} The value of $\sigma/T$  on the phase boundaries including the chiral corssover and first-order phase transitions  in the PNJL model.	}
%	\end{center}
\end{figure}

Fig.~\ref{fig:17} displays the value of $\sigma/T$ on the crossover transition line and the first-order phase transition boundaries. Along the chiral crossover phase transition line, $\sigma/T$ decreases monotonically as temperature drops. At the first-order transition, $\sigma/T$ is consistently larger in the chiral restored (high-density) phase than in the broken (low-density) phase. This trend differs from that of $\eta/s$ near the critical region in the first-order transition.

\begin{figure}[htbp]
	\centering
	\includegraphics[scale=0.37]{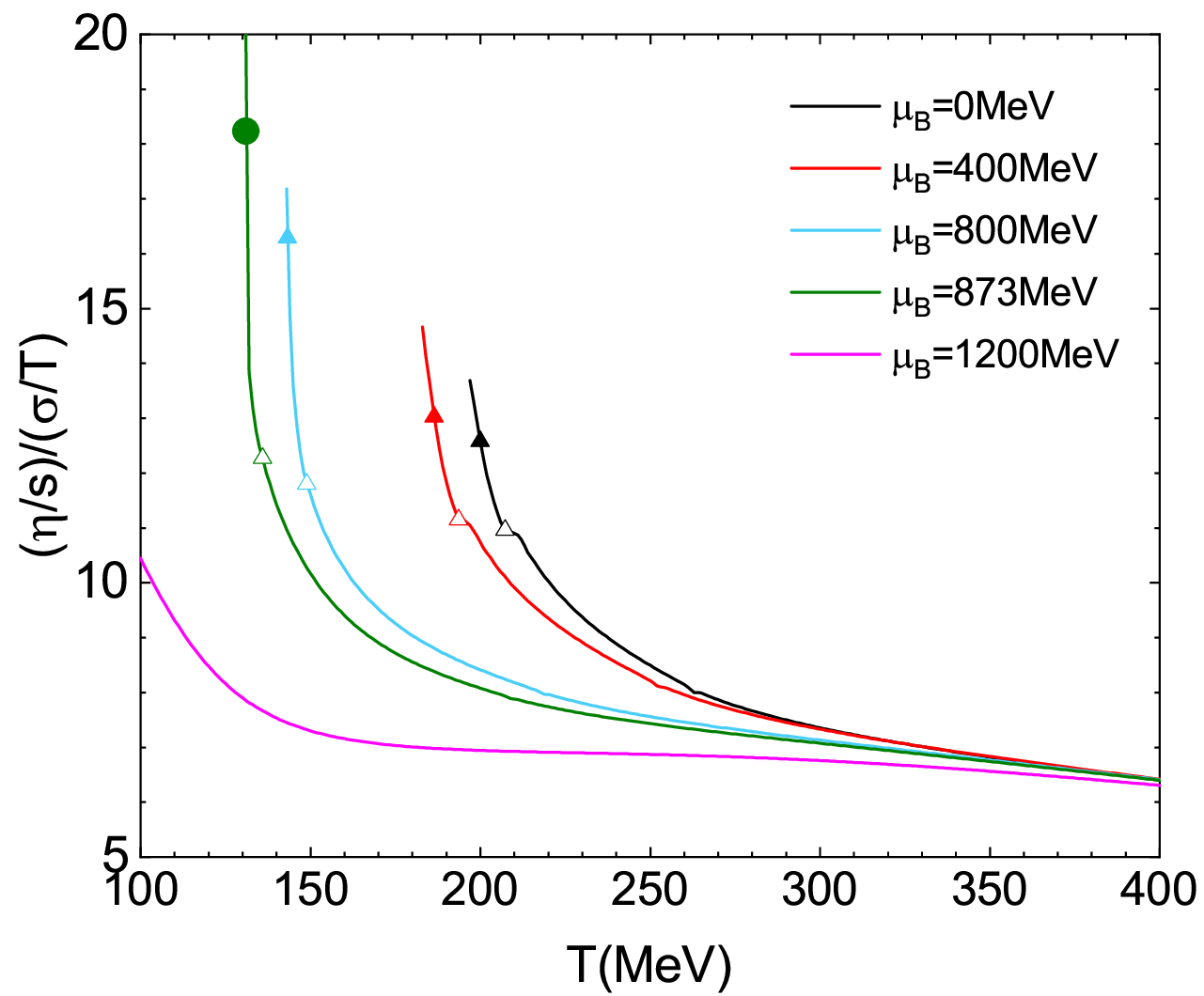}
	\caption{The ratio of $\eta/s$ to $\sigma/T$ as a function of temperature at different chemical potentials in the PNJL model. The dot, solid triangles and hollow triangles represent the value at the points where the  $\mu_B(T)$ lines intersect the CEP, chiral crossover phase transition line and pion Mott transition line, respectively.}
	\label{fig:18} 
\end{figure}

It is also interesting to calculate the dimensionless ratio of  shear viscosity to electric conductivity, an interesting quantity which possibly has certain universal features~\cite{Chakrabarti2010xy,Jain2010ip,Jain2009bi,Jaiswal2015mxa}.
We present in Fig.~\ref{fig:18} the ratio of $\eta/s$ to $\sigma/T$ as functions of temperature at different chemical potentials. 
It can be observed that the $(\eta/s)/(\sigma/T)$ increase rapidly with decreasing temperature near the Mott and chiral crossover phase transitions. Its value on the chiral crossover phase transition line ranges from approximately 12 to 18, which is consistent with the calculation in Ref.~\cite{Puglisi2014pda}.
Different from the non-monotonic behaviors of $\eta/s$ and $\sigma/T$ with temperature in Fig.~\ref{fig:6}    and  Fig.~\ref{fig:14}, $(\eta/s)/(\sigma/T)$ for each chemical potential always decreases monotonically with growing temperature. The value of $(\eta/s)/(\sigma/T)$  approaches 6.4 in the high-temperature regime. A similar result was derived in Ref.~\cite{Thakur2017hfc}. This behavior can be approximately understood as follows: at extremely high temperatures, with the chiral symmetry restoration of $u,d,s$ quarks, the differences between quark flavors (apart from the bare quark masses) become minimal. The system gradually approaches a state resembling a non-interacting ideal gas. In such a system, scale invariance dominates in the high-temperature limit, leading to a stabilization of the dimensionless $(\eta/s)/(\sigma/T)$.

\section{summary}

In this work, we explored the shear viscosity and electric conductivity of quark matter at finite temperature and density within the framework of kinetic theory with the relaxation time approximation. The flavor dependent relaxation time is derived according to the $2 \rightarrow 2 $ elastic scattering between quasiparticles including the $u,d, s$ quarks and their antiparticles. The temperature and chemical potential dependent masses of quasiparticles   are calculated in the 2+1 flavor PNJL model. The masses of exchanged mesons are derived with the random phase approximation.

We focus on investigating the characteristic behaviors of shear viscosity and electrical conductivity across the Mott transition, chiral crossover transformation, and first-order phase transition associated with a spinodal structure. It is found that the relaxation time is critical in the description of  shear viscosity and electrical conductivity.
The numerical results show that, at small chemical potential,  the relaxation time for each species of quark (anti-quark)  has a minimum near the meson Mott dissociation, and increases rapidly below the chiral crossover phase transition line. Correspondingly, both the $\eta/s$ and $\sigma/T$ exhibit a concave behavior near the pion Mott transition, and increase rapidly at the crossover phase transition with the decrease of temperature. The minimum value of $\eta/s$ at vanishing chemical potential is about 0.1, close to the KSS limit and the value extracted from the experimental data. At the CEP, the value of $\eta/s$ is above 1  within the mean-field approximation.

The calculation also indicates that, at large chemical potential (high baryon density), the behavior of $\eta/s$ in the QGP phase is mainly dominated by temperature. The value of $\eta/s$ of dense QGP is greatly enhanced with the decline of temperature. At intermediate temperature and chemical potential, the values  of  $\eta/s$ are influenced by the competition between the temperature, density effect and QCD phase transition. 

Generally, the $\sigma/T$ exhibits characteristics similar to those of $\eta/s$ in the QCD phase diagram. Our calculations further show that the dimensionless ratio $(\eta/s)/(\sigma/T)$ decreases monotonically with temperature and  approaches 6.4 in the high-temperature limit. This asymptotic value reveals that quark matter in the PNJL model resembles a non-interacting ideal gas in the high-temperature limit and exhibits approximate scale invariance.

In the present study, the calculations are performed within the mean-field approximation. A more accurate description of shear viscosity and electrical conductivity in the critical region would necessitate calculations beyond the mean field and incorporate critical dynamical effects.

\begin{acknowledgements} 
This work is supported by the National Natural Science Foundation of China under
Grant No. 12475145  and Natural Science Basic Research Plan in Shaanxi Province
of China (Program No. 2024JC-YBMS-018).
\end{acknowledgements}

%\appendix
%\section{}
%\section{Hi}
%\bibliographystyle{plain}  %根据文献顺序
%\bibliographystyle{unsrt}  %根据引用顺序
\bibliography{ref_temp.bib}
%\nocite{*}  %显示未引用文献
%\end{CJK*}
\end{document}